\documentclass{article}  
\usepackage{epsfig}
\begin{document}

\title{Zeno meets modern science}  
\author{Z.K. Silagadze \\ Budker Institute of Nuclear Physics, 
        630 090, Novosibirsk, Russia}

\date{}

\maketitle  
              	
\begin{abstract}
``No one has ever touched Zeno without refuting him''. We will not refute 
Zeno in this paper. Instead we review some unexpected encounters of Zeno with
modern science. The paper begins with a brief biography of Zeno of Elea 
followed by his famous paradoxes of motion. Reflections on continuity of 
space and time lead us to Banach and Tarski and to their celebrated paradox,
which is in fact not a paradox at all but a strict mathematical theorem,
although very counterintuitive. Quantum mechanics brings another flavour in
Zeno paradoxes. Quantum Zeno and anti-Zeno effects are really paradoxical
but now experimental facts. Then we discuss supertasks and bifurcated 
supertasks. The concept of localization leads us to Newton and Wigner and to
interesting phenomenon of quantum revivals. At last we note that the 
paradoxical idea of timeless universe, defended by Zeno and Parmenides 
at ancient times, is still alive in quantum gravity. The list of references 
that follows is necessarily incomplete but we hope it will assist interested 
reader to fill in details. 
  
\end{abstract}
	
\section{Introduction}
Concepts of localization, motion and change seems so familiar to our 
classical intuition:
everything happens in some place and everything moves from one place to
another in everyday life. Nevertheless it becomes a rather thorny issue
then subjected to critical analysis as witnessed long ago by Zeno's
paradoxes of motion. One can superficially think that the resolution of the
paradoxes was provided by calculus centuries ago by pointing out now the
trivial fact that an infinite series can have a finite sum. But on the
second thought we realize that this "resolution" assumes infinite
divisibility of space and time and we still do not know whether the physical
reality still corresponds to the continuous space and time at very small
(Plankian) scales. Even in pure mathematics the infinite divisibility leads
to paradoxical results like Banach-Tarski paradox which are hard to swallow
despite their irrefutable mathematical correctness.

More subtle under the surface truth about Zeno's paradoxes is that even if one
assumes the infinitely divisible space and time calculus does not really
resolves the paradoxes but instead makes them even more paradoxical and leads
to conclusion that things cannot be localized arbitrarily sharply. Of course
the latter is just what we expect from basic principles of quantum mechanics
and special relativity. But it is certainly amazing to find roots of these
pillars of the modern physics at Zeno's times!

\section{Zeno of Elea}
``No one has ever touched Zeno without refuting him, and every century thinks 
it worthwhile to refute him'' \cite{1}. Therefore it seems that refuting Zeno 
is eternal and unchanging affair in complete accord with the Eleatic 
philosophy. According to this philosophy all appearances of multiplicity, 
change, and motion are mere illusions. Interestingly the foundation of the 
Eleatic philosophical school was preceded by turbulent events in drastic  
contrast with its teaching of the unique, eternal, and unchanging universe
\cite{2}. The school was founded by Xenophanes (born circa 570 BC), 
a wandering exile from his native city of Colophone in Ionia. Before finally   
joining the colony at Elea, he lived in Sicily and then in Catana. The Elea
colony itself was founded by a group of Ionian Greeks which seize the site 
from the native Oenotrians. Earlier these Ionian Greeks were expelled from 
their native city of Phocaea by an invading Persian army. Having lost their 
homes, they sailed to the Corsica island and invaded it after a awful  
sea battle with the Carthaginians and Etruscans, just to drive once again into 
the sea as refugees after ten years later (in 545 BC) their rivals regained 
the island. We can just wonder about psychological influence of these events 
on the Eleatic school's belief in permanent and unalterable universe\cite{2}.
   
Zeno himself had experienced all treacherous vicissitudes of life. Diogenes 
Laertius describes him \cite{3} as the very courageous man:

``He, wishing to put an end to the power of Nearches, the tyrant (some, 
however, call the tyrant Diomedon), was arrested, as we are informed by 
Heraclides, in his abridgment of Satyrus. And when he was examined, as to his 
accomplices, and as to the arms which he was taking to Lipara, he named all 
the friends of the tyrant as his accomplices, wishing to make him feel himself 
alone. And then, after he had mentioned some names, he said that he wished to 
whisper something privately to the tyrant; and when he came near him he bit 
him, and would not leave his hold till he was stabbed. And the same thing 
happened to Aristogiton, the tyrant slayer. But Demetrius, in his treatise on 
People of the same Name, says that it was his nose that he bit off.

Moreover, Antisthenes, in his Successions, says that after he had given him 
information against his friends, he was asked by the tyrant if there was any 
one else. And he replied, "Yes, you, the destruction of the city." And that he 
also said to the bystanders, "I marvel at your cowardice, if you submit to be 
slaves to the tyrant out of fear of such pains as I am now enduring." And at 
last he bit off his tongue and spit it at him; and the citizens immediately 
rushed forward, and slew the tyrant with stones. And this is the account that 
is given by almost every one''. 

Although this account of Zeno's heroic deeds and torture at the hands of the 
tyrant is generally considered as unreliable \cite{4,5,6}, Zeno after all is 
famous not for his brevity but for his paradoxes \cite{7,8,9,10,11}.

\section{Zeno's paradoxes of motion}
Zeno was a disciple of Parmenides, the most illustrious representative of the 
Eleatic philosophy. According to Parmenides, many things taken for granted, 
such as motion, change, and plurality, are simply illusions and the  
reality is in fact an absolute, unchanging oneness. Of course, nothing 
contradicts more to our common sense experience than this belief. It is not
surprising, therefore, that Parmenides' views were ridiculed by contemporaries
(and not only). In his ``youthful effort'' Zeno elaborated a number of 
paradoxes in order to defend the system of Parmenides and attack the common 
conceptions of things. The four most famous of these paradoxes deny the 
reality of motion. The Dichotomy paradox, for example, states that it is 
impossible to cover any distance \cite{2}:

\begin{itemize}
\item
{\it There is no motion, because that which is moved must arrive at the 
middle before it arrives at the end, and so on ad infinitum.}
\end{itemize}

According to Simplicius, Diogenes the Cynic after hearing this argument from
Zeno's followers silently stood up and walked, so pointing out that it is a 
matter of the most common experience that things in fact do move \cite{11}. 
This answer, very clever and effective perhaps, is unfortunately completely
misleading, because it is not the apparent motion what Zeno questions but
how this motion is logically possible. And the Diogenes's answer does not 
enlighten us at all in this respect \cite{12}:
\begin{verbatim}
    A bearded sage once said that there's no motion.
    His silent colleague simply strolled before him, --
    How could he answer better?! -- all adored him!
    And praised his wise reply with great devotion.
    But men, this is enchanting! -- let me interject,
    For me, another grand occurrence comes to play:
    The sun rotates around us every single day,
    And yet, the headstrong Galileo was correct.
\end{verbatim}

But what is paradoxical in Zeno's arguments? Let us take a closer look. He
says that any movement can be subdivided into infinite number of ever 
decreasing steps. This is not by itself paradoxical, if we assume infinite
divisibility of space and time. What is paradoxical is an ability to perform
infinite number of subtasks in a finite time -- to perform a supertask. Any
movement seems to be a supertask according to Zeno and it is by no means
obvious that it is ever possible to perform infinite number of actions in 
a finite time. Our intuition tells us just the contrary -- that it is a clear
impossibility for finite beings to manage any supertask. In the case of  
Dichotomy it is even not clear how the movement can begin at all because there
is no first step to be taken.

Aristotle tried to resolve this situation by distinguishing potential and
actual infinities \cite{13}: ``To the question whether it is possible to pass 
through an infinite number of units either of time or of distance we must 
reply that in a sense it is and in a sense it is not. If the units are actual,
it is not possible; if they are potential, it is possible''. But Aristotle's
answer is not much better than Diogenes'. It is incomplete. In fact doubly
incomplete. According to it Zeno's infinite subdivision of a motion is purely
mathematical, just an action of imagination. But even if we accept Aristotle's
position it is desirable to show that in mathematics we have tools to handle
infinities in a logically coherent way. In fact no such tools were at 
Aristotle's disposal and they were only germinated after two thousand years 
when the notion of limit emerged, Calculus was developed and Georg Cantor 
created his set theory. We can say that Dichotomy is not mathematically 
paradoxical today. Either classical or non-standard \cite{14,15} analysis can 
supply sufficient machinery to deal with both the Dichotomy and its more 
famous counterpart, the Achilles and the tortoise paradox \cite{2}:
\begin{itemize}
\item
{\it The slower will never be overtaken by the quicker, for that which is 
pursuing must first reach the point from which that which is fleeing started, 
so that the slower must always be some distance ahead.}
\end{itemize}
This latter paradox is more impressively formulated in terms of two bodies
but in fact it is a symmetric counterpart of the Dichotomy and has a variant
involving only one moving body \cite{16}: ``To reach a given point, a body 
in motion must first traverse half of the distance, then half of what remains,
half of this latter, and so on ad infinitum, and again the goal can never be 
reached''. Therefore if the Dichotomy wonders how the motion can begin as
there is no first step, the Achilles makes it equally problematic the end of
the motion because there is no last step. Modern mathematics partly completes 
Aristotle's argument and provides a coherent mathematical picture of motion.
But all this mathematical developments, although very wonderful, do not 
answer the main question implicit in Aristotle's rebuttal of Zeno: how the 
real motion actually takes place and whether its present day mathematical 
image still corresponds to reality at the most fundamental level.

\section{Zeno meets Banach and Tarski}
Let us take, for example, infinite divisibility of space and time. This 
infinite divisibility is in fact paradoxical, even though the modern 
mathematics have no trouble to deal with this infinite divisibility. Let us 
explain what we have in mind.

Zeno's argument shows that any spatial or temporal interval contains 
uncountably many points. Nevertheless a moving body manages to traverse all 
these points in a finite time. Let us consider any division of the interval
into non-empty pairwise mutually exclusive subintervals (that is any pair of 
them have no common points). Then there exists at least one set N that 
contains one and only one point from each of the subintervals. indeed, a 
moving point body enters into a given subinterval sooner or later while 
traversing the initial interval and will remain into this subinterval for some
amount of time. We can take any point the moving body occupies during this 
time interval as an element of N. All this seems very natural and 
self-evident, and so does its natural generalization, the axiom of choice
\cite{17}:  
\begin{itemize}
\item
{\it If M is any collection of pairwise mutually exclusive, non-empty sets P,
there exists at least one set N that contains one and only one element from
each of the sets P of the collection M.}
\end{itemize}
\noindent If one may choose an element from each of the sets P of M, the set
N can evidently be formed -- hence the name of the axiom.

Now this innocently looking ``self-evident'' axiom leads to the most 
paradoxical result in the mathematics, the Banach-Tarski theorem, which is so
contrary to our intuition that is better known as the Banach-Tarski paradox. 
The most artistic presentation of this paradox can be found in the Bible 
\cite{18}:
``As he went ashore he saw a great throng; and he had compassion on them, and 
healed their sick. When it was evening, the disciples came to him and said: 
"This is a lonely place, and the day is now over; send the crowds away to go 
into the villages and buy food for themselves". Jesus said: "They need not go 
away; you give them something to eat". They said to him: "We have only five 
loaves here and two fish". And he said: "Bring them here to me". Then he 
ordered the crowds to sit down on the grass; and taking the five loaves and 
the two fish he looked up to heaven, and blessed, and broke and gave the 
loaves to the disciples, and the disciples gave them to the crowds. And they 
all ate and were satisfied. And they took up twelve baskets full of the broken
pieces left over. And those who ate were about five thousand men, besides women
and children''.

In the more formal language the Banach-Tarski theorem states that \cite{17} 
\begin{itemize}
\item
{\it in any euclidean space of dimension $n>2$, any two arbitrary bounded sets 
are equivalent by finite decomposition provided they contain interior points.}
\end{itemize}
\noindent This theorem opens, for example, the door to the following 
mathematical alchemy \cite{17}: a ball of the size of orange can be
divided into a finite number of pieces which can be reassembled by using 
merely translations and rotations to yield a solid ball whose diameter is 
larger than the size of the solar system. Of course a real orange cannot be
chopped in such a way because atoms and molecules constitute a limit of 
divisibility of any chemical substance and the pieces required in the 
Banach-Tarski theorem are so irregular that they are nonmeasurable and the 
concept of volume (Lebesgue measure) does not make sense for them. 
But does space-time itself also have a limit of divisibility? It is yet an 
open question.

The comprehensive discussion of the Banach-Tarski theorem is given in 
\cite{19} and for an elementary approach with a full proof of the theorem see
\cite{20}. One can question whether paradoxical counter-intuitive 
decompositions like the ones implied by the Banach-Tarski theorem are of any 
use in physics. Surprisingly, there were several attempts in this direction.
Pitowsky was the first (to our knowledge) to consider a certain extension of
the concept of probability to nonmeasurable sets in connection with the
Einstein-Podolsky-Rosen paradox and Bell's inequalities \cite{21,22}. Another
examples can be traced trough \cite{23,24}.

One cannot blame the axiom of choice as the only culprit of such paradoxical 
mathematical results. Even without the use of this axiom one can argue that
there is some truth in the proverb that the world is small, because the 
results proved in \cite{25} entirely constructively, without the axiom of 
choice, imply that there is a finite collection of disjoint open subsets of 
the sun that fill the whole sun without holes of positive radius and that
nevertheless can be rearranged by rigid motions to fit inside a pea and remain
disjoint. 

Maybe the the Banach-Tarski theorem and analogous paradoxical decompositions
will appear a bit less paradoxical if we realize that the Achilles and the 
tortoise paradox illustrates that any two intervals contain the same number of 
points regardless their length. Indeed, during their race Achilles and the
tortoise cover desperately different intervals. Nevertheless one can arrange
a one-to-one correspondence between points of these intervals because for
every point A from the Achilles' track there is only one point B on the track
of tortoise which the tortoise occupied at the same instant of time when 
Achilles occupied A. In fact, as Cantor proved in 1877, there is a one-to-one
correspondence of points on the interval [0, 1] and points in a square, or 
points in any n-dimensional space. Cantor himself was surprised at his own 
discovery and wrote to Dedekind \cite{26} ``I see it, but I don't believe 
it!'' 

\section{Zeno meets quantum mechanics}
Despite some paradoxical flavour, the infinite divisibility of real space-time,
although unwarranted at yet, is mathematically coherent. But Zeno's paradoxes
contain some other physical premises also that deserve careful consideration.
The Achilles and the tortoise paradox, for example, assumes some observation
procedure:
\begin{itemize}
\item check the positions of the contenders in the race.
\item check again when Achilles reach the position the tortoise occupied
at previous step.
\item repeat the previous instruction until Achilles catch the tortoise
(and this is an infinite loop because he never does).
\end{itemize}
\noindent Calculus teach us that the above observational process covers only
finite interval of time in spite of its infinitely many steps. And during
this time interval the tortoise will be indeed always ahead of Achilles.
The observational procedure Zeno is offering simply does not allow us to check
the contenders positions later when Achilles overtake the tortoise. So the 
paradox is solved? Not at all. Zeno's procedure implicitly assumes an ability
to perform position measurements. Therefore two questions remain: whether it
is possible to perform infinitely frequent measurements taken for granted by
Zeno, and how the race will be effected by back-reaction from these 
measurements. The world is quantum mechanical after all and the measurement
process is rather subtle thing in quantum mechanics.   

Simple arguments \cite{27} show that something interesting is going on if the
observational procedure of Zeno is considered from the quantum mechanical 
perspective.
Let $|\Phi,0>$ denote the initial state vector of the system (Achilles and 
the tortoise in uniform motions). After a short time $t$ the state vector will
evolve into $$|\Phi,t>=\exp{\left (-\frac{i}{\hbar}Ht\right )}|\Phi,0>\approx 
\left (1-\frac{i}{\hbar}Ht -\frac{1}{2\hbar^2} H^2t^2\right )|\Phi,0>\; ,$$ 
where $H$ is the Hamiltonian of the system assumed to be time independent. 
If now the position measurements of the competitors are performed we find that 
the initial state have still not changed with the probability
$$|<\Phi,0|\Phi,t>|^2\approx 1-\frac{(\Delta E)^2}{\hbar^2}t^2\; ,$$ where
$$(\Delta E)^2=<\Phi,0|H^2|\Phi,0>-<\Phi,0|H|\Phi,0>^2$$ is positive (there 
should be some energy spread in the initial state because we assume good 
enough localizations for Achilles and the tortoise). If these measurements 
are performed $n$-times, at intervals $t/n$, there is a probability
$$\left ( 1-\frac{(\Delta E)^2}{\hbar^2}\frac{t^2}{n^2}\right )^n$$ that at 
all times the system will be found in the initial state. But this probability
tends to unity when $n\rightarrow \infty$, because in this limit
$$\ln{\left ( 1-\frac{(\Delta E)^2}{\hbar^2}\frac{t^2}{n^2}\right )^n}\approx
-\frac{(\Delta E)^2}{\hbar^2}\frac{t^2}{n}\rightarrow 0.$$
Therefore, if the observations are infinitely frequent the initial state does
not change at all. Zeno was right after all: Achilles will never catch 
the tortoise under proposed observational scheme! This scheme implicitly 
assumes a continues monitoring of Achilles' position and therefore he
will fail even to start the race.

\begin{figure}[htb]
 \begin{center}
    \mbox{\epsfig{figure=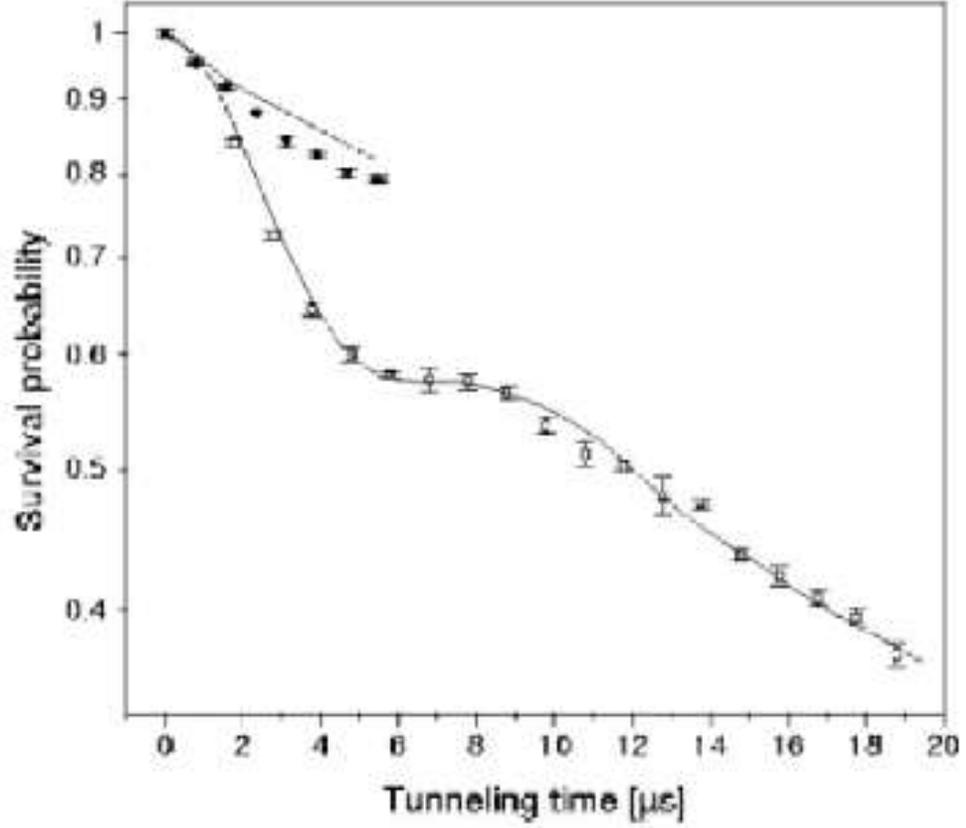}}
  \end{center}
\caption {Observation of the Quantum Zeno Effect in the experiment \cite{33}.}
\label{qz}
\end{figure}

Matters are not as simple however. Repeated measurement of a system effects 
its dynamics much more complex and delicate way than just slowing the 
evolution \cite{28}. The above described Quantum
Zeno Effect became popular after seminal paper of Misra and Sudarshan 
\cite{29}, although it dates back to Alan Turing (see \cite{30,31} and 
references wherein) and was known earlier as ``Turing's paradox''. The initial
time $t^2$ dependence of quantum mechanical evolution, from which the Quantum
Zeno Effect follows most simply, is quite general though not universal. The
experimental difficulty resides in the fact that the $t^2$ dependence takes
place usually at very short times for natural unstable systems. For example,
the "Zeno" time of the 2P-1S transition of the hydrogen atom is estimated
to be approximately $3.6 \cdot 10^{-15}~{\mathrm s}$ \cite{32}. 
Nevertheless modern experimental techniques enable to prepare artificial 
unstable systems with long enough Zeno time. In beautiful experiment \cite{33} 
ultra-cold sodium atoms were trapped in a periodic optical potential created
by a accelerated standing wave of light. Classically atoms can be trapped
inside the potential wells but they will escape by quantum tunneling.
The number that remain is measured as a function of duration of the 
tunneling. The results are shown in the Fig.\ref{qz}.
Hollow squares in this figure show  the probability of survival in the 
accelerated potential as a function of duration of the tunneling acceleration.
The solid line represents what is expected according to quantum mechanics
and we see a very good agreement with the experimental data. The Zeno time for 
this unstable system is of the order of about $\mu$s and during this short
time period the survival probability exhibits a $t^2$ drop. For longer times
we see a gradual transition from the $t^2$ dependence to linear $t$ dependence,
which corresponds to the usual exponential decay law. Such behaviour can be 
simply understood in the framework of time-dependent perturbation 
theory \cite{34}.
The probability of decay of some excited state $|i>$ under the influence of 
time independent small perturbation $V$ is given by the formula
\begin{equation}
Q(t)=\frac{1}{\hbar^2}\int\limits_{-\infty}^\infty |<i|V|k>|^2\rho_k\sin^2{
\left (\frac{(E_k-E_i)t}{2\hbar}\right )}\left (\frac{2\hbar}{E_k-E_i}
\right )^2 dE_k,
\label{GR}
\end{equation}
where  $|i>,\; |k>$ are eigenstates of unperturbed Hamiltonian and $\rho_k$
is the density of states $|k>$. For very short times one has clearly a $t^2$
dependence:
$$Q(t)\approx\frac{1}{\hbar^2} \left ( \int\limits_{-\infty}^\infty 
|<i|V|k>|^2 \rho_k dE_k \right )t^2.$$
For longer times, when $$\frac{(E_k-E_i)t}{2\hbar}$$ is not small,
one cannot replace the sine function by the first term of its Taylor 
expansion. However we can expect that only states with small $|E_k-E_i|$ 
contribute significantly in the integral, because if  
$$z=\frac{(E_k-E_i)t}{2\hbar}\gg 1$$ the integrand oscillates quickly. But 
then it can be assumed that $|<i|V|k>|^2$ and $\rho_k$  are constant and by 
using $$\int\limits_{-\infty}^\infty \frac {\sin^2{z}}{z^2} dz=\pi ,$$
one obtains linear $t$ dependence
$$Q(t)\approx \frac{2\pi}{\hbar}|<i|V|k>|^2\rho_k t.$$

In the experiment \cite{33} the number of atoms remaining in the potential
well after some time of tunneling was measured by suddenly interrupting the
tunneling by a period of reduced acceleration. For the reduced acceleration
tunneling was negligible and the atoms increased their velocity without being
lost out of the well. Therefore the remaining atoms and the ones having
tunneled out up to the point of interruption become separated in velocity
space enabling the experimenters to distinguish them. This measurement
of the number of remaining atoms projects the system in a new initial state
when the acceleration is switched back and the system returns to its unstable
state. The evolution must therefore start again with the non-exponential
initial segment and one expects the Zeno impeding of the evolution under
frequent measurements.

\begin{figure}[htb]
 \begin{center}
    \mbox{\epsfig{figure=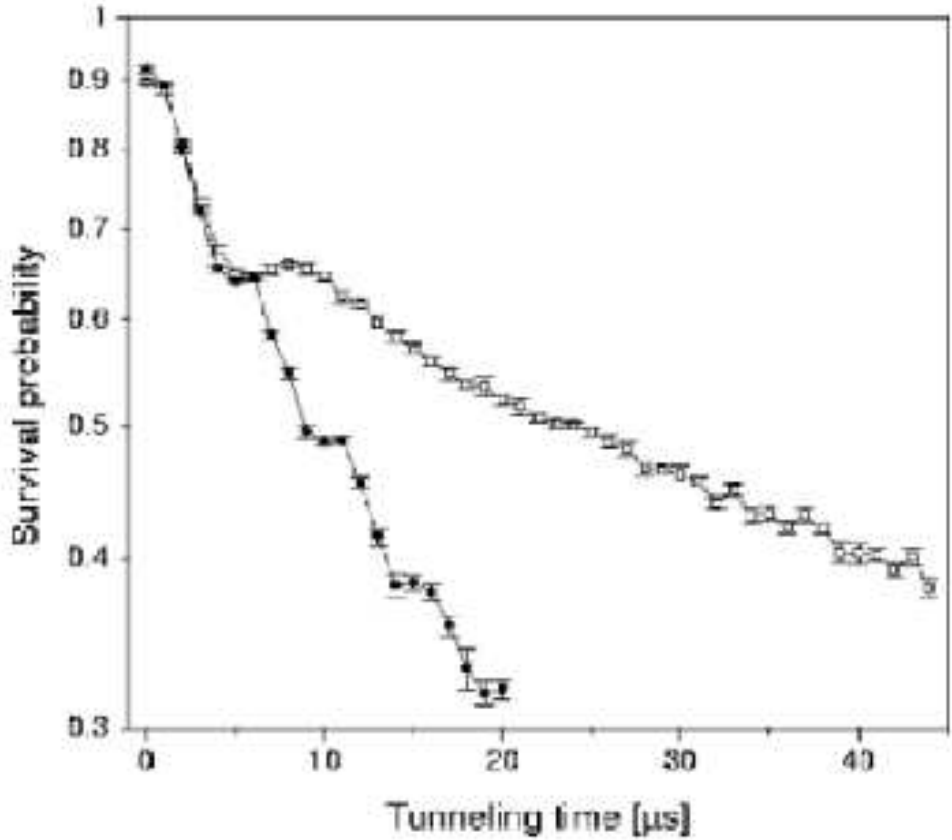}}
  \end{center}
\caption {Observation of the Quantum Anti-Zeno Effect in the experiment 
\cite{33}.}
\label{aqz}
\end{figure}
Fig.\ref{qz} really shows the Zeno effect in a rather dramatic way. The solid
circles in this figure correspond to the measurement of the survival 
probability when after each tunneling segment of 1 $\mu$s an interruption of  
50 $\mu$s duration was inserted. One clearly sees a much slower decay trend
compared to the measurements without frequent interruptions (hollow squares).
Some disagreement with the theoretical expectation depicted by the solid line
is attributed by the authors of \cite{33} to the under-estimate of the actual
tunneling time, so that in reality the decay might be slowed down even at
higher degree.

Uninterrupted decay curve shows damped oscillatory transition region between 
initial period of slow decay and the exponential decay at longer times. At
that a steep drop in the survival probability is observed immediately after 
the Zeno time. It is expected, therefore, that the decay will be not slowed
down but accelerated if frequent interruptions take place during the steep drop
period of evolution forcing the system to repeat the initial period of fast 
decay again and again after every measurement. This so called Quantum 
Anti-Zeno effect \cite{35,36,37} is experimentally demonstrated in 
Fig.\ref{aqz}. The solid circles in this figure show the evolution of the 
unstable system when after every 5 $\mu$s of tunneling the decay was 
interrupted by a slow acceleration segment. 

\section{Supertasks}
\begin{figure}[htb] 
  \begin{center}
    \mbox{\epsfig{figure=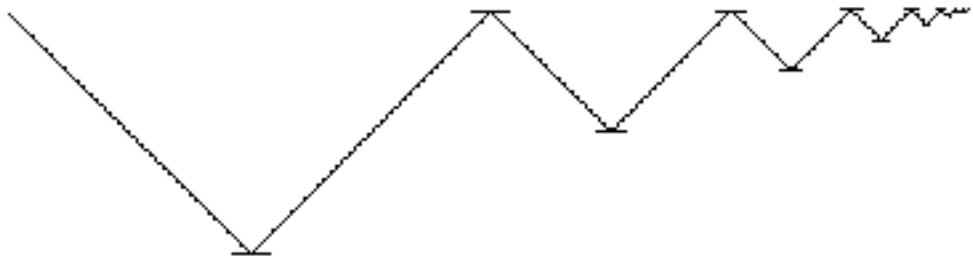}}
  \end{center}
\caption {The Zeno Zigzag.}
\label{zzg}
\end{figure}

The Zeno and Anti-Zeno effects in quantum theory are of course very 
interesting phenomena and even some practical application of the Zeno effect
in quantum computing is foreseeable \cite{38}. But in the context of Zeno 
paradoxes we are more interested in the limit of infinitely frequent 
measurements with complete inhibition of evolution. Although such limit leads
to interesting mathematics \cite{39}, its physical realizability is dubious.
The Calculus argument that it is possible for infinite sum to converge to
a finite number is not sufficient to ensure a possibility to perform a 
supertask. This becomes obvious if we somewhat modify Zeno's arguments to 
stress the role supertasks play in them. The resulting paradox, Zeno Zigzag,
goes as follows \cite{2}. A light ray  is bouncing between an infinite 
sequence of mirrors as illustrated in the Fig.\ref{zzg}.
The sizes of mirrors and their separations decrease by a factor of two on 
each step. The total length of the photon's zigzag path is finite (because
the geometric series 1+1/2+1/4+...converges), as well as the envelope box
size around the mirrors. Therefore the photon is expected to perform infinite
number of reflections in finite time and emerge on the other side of our
mirror box. But the absence of the last mirror the photon hit is obviously
troublesome now: there is no logical way for the photon to decide at what
direction to emerge from the box.

Of course, in reality it is impossible to realize the Zeno Zigzag for a number
of reasons. One cannot make arbitrarily small ``mirrors'', for example, 
because sharp localization leads to a significant momentum spread according
to uncertainty relations and then the relativity makes possible a pair 
production which will smear the ``mirror'' position. The wave-like behaviour
of the photon (or any other particle) will anyway make impossible to maintain
definite direction of the reflected photon if the mirror size is less than 
the photon wavelength.  

The question, however, naturally arises whether supertasks are logically 
impossible irrespective of the nature of physical reality which may restrict
their practical realization. To support the opinion that the very notion of
completing an infinite sequence of acts in a finite time is logically 
contradictory, Thomson suggested the following supertask \cite{40}. A lamp is
switched ON and OFF more and more rapidly so that at the end of the two 
minutes a supertask of infinite switching of the lamp is over. The question 
now is whether the lamp is in the ON state or in the OFF state after this two
minutes. Clearly the lamp must be in one of these states but both seem equally
impossible. The lamp cannot be in the ON state because we never turned it on
without immediately turning it off. But the lamp cannot be in the OFF state 
either because we never turned it off without at once turning it on. It seems
impossible to answer the question and avoid a contradiction.

\begin{figure}[htb] 
  \begin{center}
    \mbox{\epsfig{figure=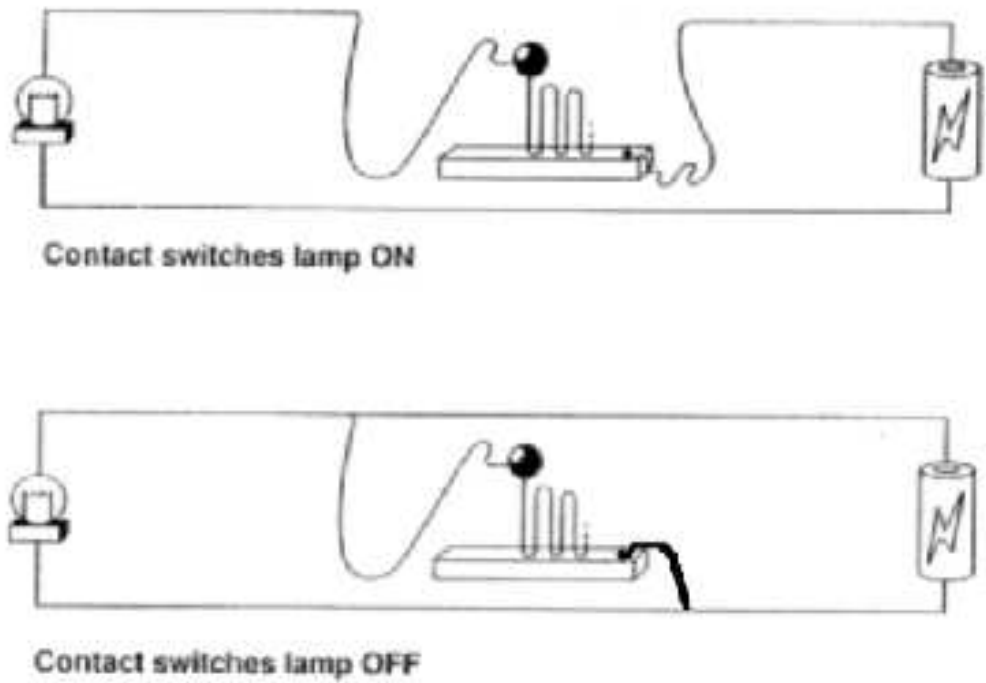}}
  \end{center}
\caption {Alternative switching mechanisms for Thomson's lamp from \cite{42}.}
\label{Tlamp}
\end{figure}
The Thomson lamp argument is seductive but fallacious \cite{41,42}. 
Surprisingly there is a coherent answer to the Thomson's question without any
contradiction. To come to this answer, it is instructive to consider another
supertask \cite{42} which is not paradoxical in any obvious way. A ball
bounces on a hard surface so that on each rebound it loses $1-k$-fraction of 
its velocity prior to the bounce, where $0<k<1$. The ball will perform 
infinitely many bounces in a finite time because, in classical mechanics,
the time between bounces is directly proportional to the initial velocity of
the ball and the geometric series $1,\;k,\;k^2,\;k^3,...$ has finite sum
$1/(1-k)$. Now let us use this bouncing ball as a switching mechanism for the
Thomson's lamp. Then it is immediately obvious that depending on the 
organization of the circuit the lamp can be in either state (ON or OFF) after
the supertask is completed, see Fig.\ref{Tlamp}.  

Logic once again demonstrates its flexibility. Note that even such a weird
notion as the lamp being in a superposition of the ON and OFF states makes
perfect sense in quantum mechanics. Although, as was indicated above, we can
make the Thomson's lamp logically consistent without any such weirdness.
But other surprises with supertasks are lurking ahead.

\begin{figure}[htb] 
  \begin{center}
    \mbox{\epsfig{figure=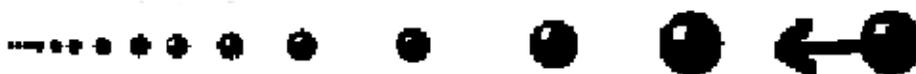}}
  \end{center}
\caption {P\'{e}rez Laraudogoitia's supertask.}
\label{Balls}
\end{figure}

In \cite{43} P\'{e}rez Laraudogoitia constructed a beautifully simple 
supertask which demonstrates some weird things even in the context of 
classical mechanics. Fig.\ref{Balls} shows an infinite set of identical
particles arranged in a straight line. The distance between  the particles
and their sizes decrease so that the whole system occupies an interval of
unit length. Some other particle of the same mass approaches the system from 
the right with unit velocity. In elastic collision with identical particles 
the velocities are exchanged after the collision. Therefore a wave of elastic 
collisions goes through the system in unit time. And what then? Any particle
of the system and the projectile particle comes to rest after colliding its
left closest neighbor. Therefore all particles are at rest after the collision
supertask is over and we are left with paradoxical conclusion \cite{43} that 
the total initial energy (and momentum) of the system of particles can 
disappear by means of an infinitely denumerable number of elastic collisions, 
in each one of which the energy (and momentum) is conserved!

If you fill uneasy about this energy-momentum nonconservation, here is the same
story in  more entertaining incarnation \cite{44}.

Suppose you have some amount of one dollar bills and the Devil approaches you
in a nefarious underground bar. He says that he has a hobby of collecting
one dollar bills of particular serial numbers and it happened that you do have
one such bill. So he is offering you a bargain: he will give you ten one 
dollar bills for this particular one. Should you accept the bargain? Why not,
it seems so profitable. You agreed and the bargain is done. After half an hour
the Devil appears again with the similar offer. Then after a quarter-hour and
so on. The time interval between his appearances decreases by a factor of two
each time. The amount of your money increases quickly. After an hour infinite
number of bargains are done and how much money will you have? You would have
a lot if the Devil wanted you to succeed. But he tries to perish your soul not
to save it and during the bargains is indeed very selective about bill serial
numbers: he always takes the bill with smallest serial number you posses and 
instead gives the bills with serial numbers greater than any your bill's 
at that moment. Under such arrangement any bill you posses will end its path
into Devil's pocket. Indeed, for any particular bill at your disposal at some 
instant you will have only finite amount of bills with less serial numbers and 
you never will get additional bills with smaller serial numbers. Therefore
sooner or later this particular bill will become the bill with smallest serial 
number among your bills and hence subject of the exchange. We come again to 
the strange conclusion that in spite of your money's continuous growth during
the transaction you will have no bill at all left after the infinite 
transaction is over! 
 
Of course, this example does not enlighten us much about how the initial energy
disappears in the P\'{e}rez Laraudogoitia's supertask. However it clearly 
shows a somewhat infernal flavour the supertasks have. And not only the 
energy-momentum conservation is on stake. Classical dynamics is time reversal
invariant. Therefore the following process, which is the time reversal of the
P\'{e}rez Laraudogoitia's supertask, is also possible \cite{43}: 
a spontaneous self-excitation can propagate through the infinite system of
balls at rest causing the first ball to be ejected with some nonzero velocity.
The system displays not only the energy-momentum non-conservation but also
indeterminism \cite{43,45,46}.

The locus of the difficulty is the same as in the Zeno paradoxes: there is no
last member in a sequence of acts (collisions) and therefore there is no last
ball to carry off the velocity \cite{47}. The supertask illustrate the 
indeterminism of classical Newtonian or even relativistic dynamics as far as
infinite localization of bodies is admissible. In quantum mechanics balls 
cannot be simultaneously at rest and infinitely localized thanks to 
uncertainty relations. Therefore P\'{e}rez Laraudogoitia's supertask will not
persist in quantum theory. However, it was shown \cite{48} that 
a (nonrelativistic) quantum mechanical supertask can be envisaged in which
the deterministic time development of the wave function is lost and 
spontaneous self-excitation of the ground state is allowed. Yet
pathologies disappear if one demands normalizability of the state vector.
In this sense quantum mechanical supertasks  are better behaved than their 
classical counterparts \cite{48}.
 
Surprisingly and  contrary to common wisdom, classical Newtonian physics is 
more hostile and unfriendly to determinism than either quantum mechanics or 
special relativity \cite{49}. Another example of esoteric behaviour of 
seemingly benign Newtonian system was given by Xia \cite{50} while solving 
the century-old intriguing problem of noncollisional singularities. Xia's
construction constitutes in fact a supertask and involves only five bodies
interacting via familiar Newtonian inverse square force law. The essentials 
of this supertask can be explained as follows \cite{51}. Imagine a system of
two equal masses $M$ moving in the $x-y$ plane under Newton's law of 
attraction with center of mass at rest, and the third mass $m\ll M$ (so that
it does not disturb the motion of the first two) on the $z$-axis which goes
through the center of mass of the planar binary system. It is clear from the 
symmetry that the total force of gravitational attraction acting on the mass 
$m$ has only $z$-component and therefore the third mass will experience a one 
dimensional motion along the $z$-axis. When the mass $m$ passes through and 
is lightly above the binary plane, we can arrange extremely powerful downward
pull on $m$ if the highly elliptical binary has its closest approach at that 
moment. In fact, assuming an ideal case of point masses, the downward force 
acting on $m$ can be made arbitrarily strong by just adjusting the separating
distances among the bodies. As a result the third mass is jolted down with
high velocity while the binary starts separating and loses significantly its
braking effect on $m$.

\begin{figure}[htb] 
  \begin{center}
    \mbox{\epsfig{figure=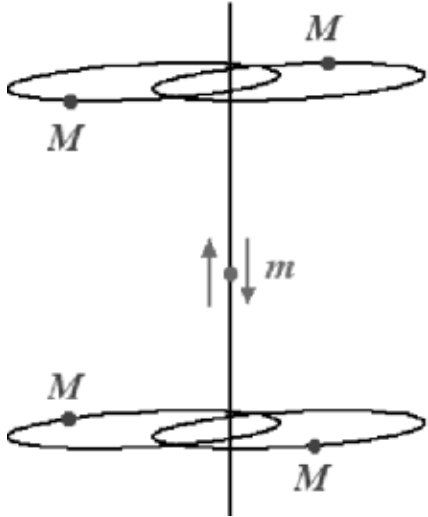}}
  \end{center}
\caption {Xia's five body supertask.}
\label{Xia}
\end{figure}

The mirror replica of the binary, see Fig.\ref{Xia}, is placed at large 
distance (not to disturb the first binary) further down to prevent the mass
$m$ from being expelled to infinity. In case of proper timing, the second
binary will break $m$'s downward motion and will thrust it upwards with even 
higher velocity.   

Xia was able to show that there exists a Cantor set of the initial conditions
allowing to repeat this behaviour infinitely often in a finite time. As 
a result the four of five bodies from the Xia's construction will escape
to spatial infinity in a finite time, while the fifth will oscillate back and
forth among the other four with ever increasing speed.

Xia's supertask does not violate energy conservation, the bodies drawing out
their energies  from the infinitely deep $1/r$ potential well. However it 
implies indeterminacy in idealized Newtonian world. The time reverse of this
supertask is an example of ``space invaders'' \cite{49}, particles appearing
from spatial infinity in a surprise attack. Note that \cite{49} ``the 
prospects for determinism brighten considerably when we leave classical 
spacetimes  for Minkowski spacetime, the spacetime setting for special 
relativistic theories''. Because with no superluminal propagation there are 
no space invaders.

Leaving aside interesting philosophical questions raised by Xia's supertask,
it seems otherwise completely artificial and far from reality. Curiously, some
ideas involved in this construction can be used by mankind, admittedly in the
very long run, to transfer orbital energy from Jupiter to the Earth by 
a suitable intermediate minor space body, causing the Earth's orbit to expand
and avoid an excessive heating from enlarging and brightening Sun at the last 
stage of its main sequence life \cite{52}.

\section{Bifurcated supertasks}
Let us return to Zeno. Hermann Weyl pointed out \cite{53} another possibility
mankind can benefit from Zeno supertask: ``if the segment of length 1 really 
consists of infinitely many subsegments of length 1/2, 1/4, 1/8, . . ., as of
``chopped-off'' wholes, then it is incompatible with the character of the 
infinite as the ``incompletable'' that Achilles should have been able to 
traverse them all. If one admits this possibility, then there is no reason 
why a machine should not be capable of completing an infinite sequence of 
distinct acts of decision within a finite amount of time; say, by supplying 
the first result after 1/2 minute, the second after another 1/4 minute, the 
third 1/8 minute later than the second, etc. In this way it would be possible,
provided the receptive power of the brain would function similarly, to 
achieve a traversal of all natural numbers and thereby a sure yes-or-no 
decision regarding any existential question about natural numbers!''

Relativity brings additional flaviour in discussion of Weyl's infinity 
machines. One can imagine, for example, bifurcated supertasks \cite{42,54}.
To check the Goldbach's conjecture whether every even number greater than two 
can be written as the sum of two primes, the Master organizes two space 
missions. In one of them a computer, the Servant, is sent with constant 
acceleration which examines all even numbers one case after the other. In 
another space mission the Master himself contrives to accelerate in such a way
to keep the Servant within his causal horizon. It is possible to arrange
Master's acceleration so that the Servant disappears from his causal shadow
only after spending infinite amount of time on the Goldbach's conjecture,
while in the Master's frame the amount of time passed remains finite. If the
Servant finds a counterexample to the Goldbach's conjecture it sends 
a message to the Master and the latter will know that the conjecture is false.
If no message is received, however, the Master will know in a finite time that
Goldbach was wright.

This looks too good to be true and indeed the realization of such Pitowsky 
\cite{55} infinite machine is suspicious for a number of reasons \cite{42,54}.
The Servant moves with constant acceleration during infinite proper time. 
Therefore it needs an infinite fuel supply. Moreover, the Master's 
acceleration increases without limit and eventually he would be crushed to
death by artificial gravity. There is a conceptual problem also. At no point
on his world line does the Master have a causal access to all events on the
Servant's world line. This means there is no Moment of Truth, if the 
Goldbach's conjecture is true, at which the Master attained that knowledge.
Emergence of the knowledge about validity of the Goldbach's conjecture will
be as mysterious in Pitowsky bifurcated supertask as is the disappearance
of energy in P\'{e}rez Laraudogoitia's supertask.

But general relativity allows to improve the above construction by admitting
the so called Malament-Hogarth spacetimes \cite{54,56}. In such a spacetime
there is a point (the Malament-Hogarth point) at Master's world line such that
entire infinite history of the Servant's world line is contained in this 
point's causal shadow. From that event on the Master will be enlightened about
validity of that particular number theoretic problem through the Servant's 
infinite labors.

Some Malament-Hogarth spacetimes seem quite reasonable physically. Among them
are such well known spacetimes as the anti-de Sitter spacetime, 
Reissner-Nordstr\"{o}m spacetime and Kerr-Newman spacetime \cite{57}, the 
latter being the natural outcome of the late-time evolution of a collapsed 
rotating star. Therefore it is not excluded that ``if the Creator had 
a taste for the bizarre we might find that we are inhabiting one of them'' 
\cite{42}. Even if this proves to be the case, the practical realization of
infinite computation is not at all guaranteed. Physics beyond the classical 
general relativity, for example proton decay and other issues related to the 
long term fate of the universe \cite{58}, can prohibit the infinite 
calculations. There is still another reason that can make a practical 
realization of bifurcated supertasks dubious \cite{55}: the Servant might 
have infinite time to accomplish its labors but not enough computation space. 
the argument goes as follows \cite{59,60}.

Any system which performs computation as an irreversible process will 
dissipate energy. Many-to-one logical operations such as AND or ERASE are not
reversible and require dissipation of at least $kT\ln{2}$ energy per bit of 
information lost when performed in a computer at temperature $T$. While 
one-to-one logical operations such as NOT are reversible and in theory can be 
performed without dissipation \cite{61}.  

That erasure of one bit information has an energy cost $kT\ln{2}$ (the 
Landauer's principle) can be demonstrated by Maxwell's demon paradox 
originally due to Leo Szilard \cite{62,63}.
\begin{figure}[htb]
 \begin{center}
    \mbox{\epsfig{figure=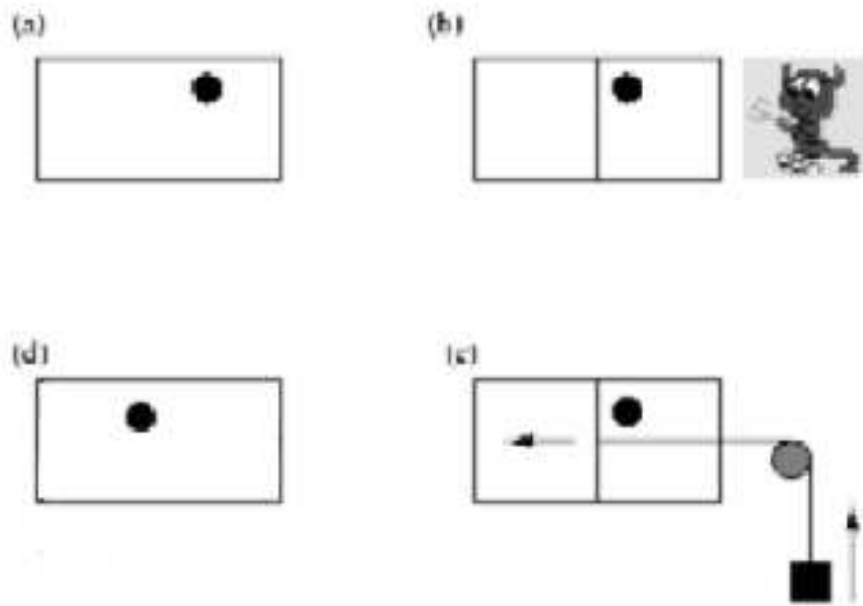}}
  \end{center}
\caption {Szilard's demon engine.}
\label{demon}
\end{figure}

A schematic view of the Szilard's demon engine is shown in the 
Fig.\ref{demon}. Initially the entire volume of a cylinder is available to 
the one-atom working gas (step (a)). At step (b) the demon measures the 
position of the atom and if it is found in the right half of the cylinder 
inserts a piston. At step (c) the one-atom gas expands isothermally by 
extracting  necessary heat from the reservoir and lifts the load. At step
(d) the piston is removed and the system is returned to its initial state ready
to repeat the cycle.

It seems the Szilard's demon engine defeats the second law of thermodynamics:
the heat bath, which has transferred energy to the gas, has lowered its 
entropy while the engine has not changed its entropy because it returned to 
the initial state. But this reasoning is fallacious because it misses an 
important point: the demon has not returned to his initial state. He still
possesses one bit information about left-right position of the atom he 
recorded. The system truly to return to its initial state it is necessary to
erase this information from the demon's memory. According to the 
Landauer's principle, one has to pay $kT\ln{2}$ energy cost for this erasure.
On the other hand work done during isothermal expansion equals
$$\int\limits_{V/2}^V pdV=\int\limits_{V/2}^V \frac{m}{\mu}RT \frac{dV}{V}=
\frac{m}{\mu}RT \ln{2}$$
and normalized to one atom this is just $(R/N_A)T \ln{2}=kT\ln{2}$. Therefore
all extracted work is paid back for erasure of the demon's memory and the net 
effect of the circle is zero. The second law defeats the demon. 
  
In principle all computations could be performed using reversible logical 
operations and hence without energy cost \cite{61}. Interestingly enough, some 
important stages of biomolecular information processing, such as 
transcription of DNA to RNA, appear to be accomplished by reversible chemical
reactions \cite{64}. Real computers, however, are subject to thermal 
fluctuations that cause errors. To perform reliable computations, therefore,
some error-correcting codes must be used to detect inevitable errors and
reject them to the environment at the cost of energy dissipation \cite{61}.

The Servant computer from bifurcated supertask needs to consume energy from
surrounding universe to perform its task. Different energy mining strategies
were considered in \cite{59} and it was shown that none of them allows to
gather infinite amount of energy even in infinite time irrespective assumed
cosmological model. To perform infinite number of computations with finite
available energy the Servant should be able to continuously decrease its
operating temperature to reduce the energy cost of computations. But the 
Servant is not completely free to choose its temperature: the waste heat 
produced while performing computations must be radiated away to avoid 
overheating. But physical laws place limits on the rate at which the waste 
heat can be radiated. Assuming that the electromagnetic dipole radiation by 
electrons is the most efficient way to get rid of the heat, Dyson argued
\cite{65} that there is a lower limit on the operating temperature
$$T>\frac{Q}{N_e}10^{-12}K,$$
where $Q$ is a measure of the complexity of the computing device ($Q\sim
10^{23}$ for humans) and $N_e$ is the number of electrons in the computer.
Since $Q/N_e$ cannot be made arbitrarily small, it seems infinite
computations are impossible.

But where is a way out \cite{65}: hibernation. The Servant can compute 
intermittently while continuing to radiate waste heat into space during its 
periods of hibernation. This strategy will allow to operate below the Dyson
limiting temperature. But eventually the wavelength of thermal radiation will 
become very large compared to the characteristic size of the computer, the 
thermal energies will be small compared to the characteristic quantized 
energy levels of the system and radiation will be suppressed by an exponential
factor compared to the estimates of Dyson \cite{59}. Therefore the Servant 
must increase its size as time goes by. It also needs more and more memory to
store digital codes of ever increasing even and prime numbers. This is another
reason it to increase in size. But as was mentioned above there is only finite
amount of energy, and hence material, available. Therefore it appears the 
servant will not be able to accomplish its infinite labors. 

\section{Zeno meets Newton and Wigner}
Our discussion of Zeno's paradoxes indicates a rather subtle role the concept
of localization plays in description of motion.
It is not surprising therefore that the notion of localization in
relativistic quantum mechanics was intensively examined. Many concepts of
localization have been proposed but we focus on the one known as the
Newton-Wigner localization \cite{66,67}. 
Initially this concept of localization and the
corresponding position operators were suggested in the context of the single
particle relativistic quantum mechanics. But later their result was
reformulated in quantum field theory also, where the concept of local
observables is the central concept leading to numerous troubles (infinities)
which are swept under carpet with great artistic skill by renormalization.
Many troubles related to localization in relativistic quantum field theory have
their formal root in the Reeh-Schlieder theorem which in the non-formal
artistic formulation of Hans Halvorson \cite{68} 
states that the vacuum is  seething
with activity  at the local level: any local event has the nonzero
probability to occur in the vacuum state for the standard localization scheme.
The Newton-Wigner localization scheme avoids some consequences of the
Reeh-Schlieder theorem and leads to a mathematical structure which seems
more comfortable for our a priori physical intuition about localization 
\cite{69}. But the story is not yet over. The suggested generalizations of the
Newton-Wigner localization are still not immune against the full strength of
the Reeh-Schlieder theorem and have their own counterintuitive features
\cite{68,70}.

Let us sketch the derivation of the Newton-Wigner position operator for Dirac
particles \cite{71}. The Newton-Wigner operator $\hat Q^k$ can be defined as
the operator whose eigenvalues are the most localized wave-packets formed
from only positive-energy solutions of the Dirac equation. Let $\psi^{(s)}
_{(\vec{y})}(\vec{x})$ be such a wave-function describing an electron with 
spin projection $s$ localized at the point $\vec{y}$ at the time $t=0$. 
Defining the scalar product by
$$(\psi,\phi)=\int \psi^+(\vec{x})\phi(\vec{x})\; d\vec{x}=
\int \psi^+(\vec{p})\phi(\vec{p})\; d\vec{p},$$
the natural normalization condition for these states is
\begin{equation}
\left (\psi^{(s)}_{(\vec{y})},\psi^{(r)}_{(\vec{z})}\right )=
\delta_{rs}\delta(\vec{z}-\vec{y}).
\label{norm}
\end{equation}
Translational invariance implies
$$\psi^{(s)}_{(\vec{y})}(\vec{x})=\psi^{(s)}_{(\vec{y}+\vec{a})}(\vec{x}+
\vec{a}),$$
or in the momentum space
\begin{equation}
\psi^{(s)}_{(\vec{y})}(\vec{p})=e^{i\vec{p}\cdot \vec{a}}
\psi^{(s)}_{(\vec{y}+\vec{a})}(\vec{p}),
\label{trans}
\end{equation}
where momentum space wave-functions are defined through
$$\psi(\vec{p})=\frac{1}{(2\pi)^{3/2}}\int \psi(\vec{x})
e^{-i\vec{p}\cdot \vec{x}}\; d\vec{x}. $$
Equations (\ref{norm}) and (\ref{trans}) imply
\begin{equation}
\psi^{(s)+}_{(\vec{y})}(\vec{p})\;\psi^{(r)}_{(\vec{y})}(\vec{p})=
\frac{\delta_{rs}}{(2\pi)^3}.
\label{eq4}
\end{equation}  
On the other hand
\begin{equation}
\psi^{(s)}_{(\vec{y})}(\vec{p})=f_{(\vec{y})}(\vec{p})\;u(\vec{p},s),
\label{eq5}
\end{equation}
where $u(\vec{p},1/2)$ and $u(\vec{p},-1/2)$ are two independent 
positive-energy solutions of the Dirac equation, which can be
taken in the form
$$u(\vec{p},s)=\Lambda_+(\vec{p})\;u(\vec{0},s),$$
where the rest state four-component spinors are
$$u(\vec{0},1/2)=(1,0,0,0)^T,\;\;\; u(\vec{0},-1/2)=(0,1,0,0)^T,$$
and $\Lambda_+(\vec{p})$ is a positive-energy projection operator
for a Dirac particle of mass $m$:
$$\Lambda_+(\vec{p})=\frac{1}{2}\left ( 1+\frac{\vec{\alpha}\cdot\vec{p}+
\beta m}{p_0}\right)=\frac{(\hat p+m)\gamma_0}{2p_0},\;\;p_0=\sqrt{\vec{p}^2+
m^2}.$$
The normalization of $u(\vec{p},s)$ is
\begin{equation}
u^{+}(\vec{p},s)u(\vec{p},s^\prime)=u^{+}(\vec{0},s)\Lambda_+(\vec{p})
u(\vec{0},s^\prime)=\frac{p_0+m}{2p_0}\delta_{ss^\prime}.
\label{eq6}
\end{equation}
From equations (\ref{eq4}-\ref{eq6}) we get
$$|f_{(\vec{y})}(\vec{p})|^2=(2\pi)^{-3}\frac{2p_0}{p_0+m}.$$
We fix the phase of $f_{(\vec{y})}(\vec{p})$ by assuming
$$f^*_{(\vec{0})}(\vec{p})=f_{(\vec{0})}(\vec{p}).$$
Then (\ref{trans}) indicates that the momentum space wave-function for the 
electron localized at $\vec{y}$ at the time $t=0$ has the form 
\begin{equation}
\psi^{(s)}_{(\vec{y})}(\vec{p})=\frac{1}{(2\pi)^{3/2}}\sqrt{\frac{2p_0}{p_0+m}}
e^{-i\vec{p}\cdot\vec{y}}\;u(\vec{p},s).
\label{eq7}
\end{equation}
Now we construct the Newton-Wigner position operator $\hat Q^k$ for which 
(\ref{eq7}) is the eigenfunction with eigenvalue $y^k$:
$$\hat Q^k\psi^{(s)}_{(\vec{y})}(\vec{p})=y^k \psi^{(s)}_{(\vec{y})}
(\vec{p}).$$
Assuming that $\psi^{(s)}_{(\vec{y})}$ eigenfunctions form a complete system 
for positive-energy solutions, we get for any positive-energy wave-function
$\psi (\vec{p})$
$$\hat Q^k\psi (\vec{p})=\hat Q^k \sum_s\int d\vec{y} \left (\psi^{(s)}_
{(\vec{y})},\psi\right)\psi^{(s)}_{(\vec{y})}(\vec{p})=
\sum_s\int d\vec{y} y^k \left (\psi^{(s)}_{(\vec{y})},\psi\right)\psi^{(s)}_
{(\vec{y})}(\vec{p}). $$
Substituting (\ref{eq7}) and performing $y$-integration we get
$$\hat Q^k\psi (\vec{p})=\int d\vec{q} \frac{2\sqrt{q_0p_0}}{(q_0+m)(p_0+m)}
\sum_s u(\vec{p},s)\,u^+(\vec{q},s)\left( -i\frac{\partial}{\partial q^k}
\delta(\vec{q}-\vec{p})\right ) \psi(\vec{q}).$$
But
$$\sum_s u(\vec{p},s)\,u^+(\vec{q},s)=\Lambda_+(\vec{p})\left (\sum_s 
u(\vec{0},s)\,u^+(\vec{0},s)\right )\Lambda_+(\vec{q})=\frac{1}{2}
\Lambda_+(\vec{p})(1+\gamma_0)\Lambda_+(\vec{q}).$$
Therefore
$$\hat Q^k\psi (\vec{p})=\Lambda_+(\vec{p})(1+\gamma_0)\sqrt{\frac{p_0}
{p_0+m}}\left( i\frac{\partial}{\partial p^k}\right )\sqrt{\frac{p_0}
{p_0+m}}\Lambda_+(\vec{p})\psi (\vec{p}).$$
From this equation it is clear that
\begin{equation}
 \hat Q^k=\Lambda_+(\vec{p})(1+\gamma_0)\sqrt{\frac{p_0}
{p_0+m}}\left( i\frac{\partial}{\partial p^k}\right )\sqrt{\frac{p_0}
{p_0+m}}\Lambda_+(\vec{p}).
\label{eq8}
\end{equation}
Newton-Wigner position operator simplifies in the Foldy-Wouthuesen 
representation. In this representation its eigenfunctions are obtained 
via unitary transformation \cite{72,73,74}:
$$\phi^{(s)}_{(\vec{y})}(\vec{p})=e^{iW}\psi^{(s)}_{(\vec{y})}(\vec{p}),$$
where the Foldy-Wouthuesen unitary operator is
$$e^{iW}=\sqrt{\frac{2p_0}{p_0+m}}\left\{\frac{1}{2}(1+\gamma_0)
\Lambda_+(\vec{p})+\frac{1}{2}(1-\gamma_0)\Lambda_-(\vec{p})\right\}.$$
Using the identity
$$\frac{1}{2}(1+\gamma_0)\Lambda_+(\vec{p})\,u(\vec{0},s)=
\frac{1}{2}(1+\gamma_0)\Lambda_+(\vec{p})\frac{1}{2}(1+\gamma_0)
\,u(\vec{0},s)=\frac{p_0+m}{2p_0}u(\vec{0},s)$$
we get
$$\phi^{(s)}_{(\vec{y})}(\vec{p})=\frac{1}{(2\pi)^{3/2}}e^{-i\vec{p}\cdot
\vec{y}}u(\vec{0},s).$$
These are Newton-Wigner position operator eigenstates in the Foldy-Wouthuesen
representation. The corresponding position operator can be derived from them
in the above described manner. The result is (in the momentum space)
$$\hat {\vec{Q}}_{(W)}=\frac{1}{2}(1+\gamma_0)i\frac{\partial}
{\partial \vec{p}}\;,$$
or in configuration space
$$\hat {\vec{Q}}_{(W)}=\frac{1}{2}(1+\gamma_0)\vec{x}=\frac{1}{2}
(1+\gamma_0)\vec{x}\frac{1}{2}(1+\gamma_0).$$
Therefore the Newton-Wigner position operator appears to be just the 
positive-energy projection of the Foldy-Wouthuesen ``mean position operator''
\cite{72,73,74} and hence (\ref{eq8}) is equivalent to
\begin{equation}
\hat{\vec{Q}}=\Lambda_+(\vec{p})\left\{i\frac{\partial}{\partial \vec{p}}+
i\frac{\beta \vec{\alpha}}{2p_0}-\frac{i\beta(\vec{\alpha}\cdot\vec{p})\vec{p}
+(\vec{\sigma}\times\vec{p})p_0}{2p_0^2(p_0+m)}\right\}\Lambda_+(\vec{p}),
\label{FWmp}
\end{equation}
where $\vec{\sigma}=(1/2i)\vec{\alpha}\times\vec{\alpha}.$

Let us compare time evolutions of the conventional (Dirac) and 
Newton-Wigner position operators in the Heisenberg picture \cite{75,76}.
The Dirac position $\vec{x}(t)=e^{i\hat Ht}\vec{x}e^{-i\hat Ht}$ 
satisfies the Heisenberg equation of motion
$$\frac{d\vec{x}(t)}{dt}=i[\hat H,\vec{x}(t)]=e^{i\hat Ht}\vec{\alpha}
e^{-i\hat Ht}=\vec{\alpha}(t),$$
where we have used $\hat H=\vec{\alpha}\cdot\hat{\vec{p}}+\beta m$ and the 
canonical commutation relations (we are using $\hbar=c=1$ convention 
throughout this section)
$$[x_i,\hat p_j]=i\delta_{ij}.$$
On the other hand by using
$$[\alpha_i,\alpha_j]=2(\delta_{ij}-\alpha_j\alpha_i)=2(\alpha_i\alpha_j-
\delta_{ij}),\;\;[\beta,\vec\alpha]=2\beta\vec\alpha=-2\vec\alpha\beta,$$
we get
$$\frac{d\vec{\alpha}(t)}{dt}=i[\hat H,\vec{\alpha}(t)]=
ie^{i\hat Ht}[\hat H,\vec{\alpha}]e^{-i\hat Ht}=2i[\hat{\vec{p}}-\vec{\alpha}
(t)\hat H]=2i[\hat H \vec{\alpha}(t)-\hat{\vec{p}}].$$
Differentiating once more, we obtain
\begin{equation}
\frac{d}{dt}\dot{ \vec{\alpha}}(t)=-2i\dot{ \vec{\alpha}}(t)\hat H=
2i\hat H\dot{ \vec{\alpha}}(t).
\label{dadot}
\end{equation}
Therefore
$$\dot{ \vec{\alpha}}(t)\hat H=-\hat H \dot{\vec{\alpha}}(t)$$
and the solution of (\ref{dadot}) is
\begin{equation}
\dot{ \vec{\alpha}}(t)\equiv \frac{d\vec{\alpha}(t)}{dt}=
 \dot{ \vec{\alpha}}(0)e^{-2i\hat H t}=2i(\hat{\vec{p}}-\vec{\alpha}
\hat H)e^{-2i\hat H t}.
\label{adot}
\end{equation}
One can integrate (\ref{adot}) by using
$$\int\limits_0^t e^{-2i\hat H \tau} d\tau=\frac{i}{2}\hat{H}^{-1}\left (
e^{-2i\hat H t}-1\right )$$
and the result is 
$$\vec{\alpha}(t)=\hat{\vec{p}}\hat{H}^{-1}+(\vec{\alpha}-\hat{\vec{p}}
\hat{H}^{-1})e^{-2i\hat H t}.$$
Inserting this into
$$\frac{d\vec{x}(t)}{dt}=\vec{\alpha}(t)$$
and  integrating the resulting equation, we get the Dirac position operator
in the Heisenberg picture
\begin{equation}
\vec{x}(t)=\vec{x}+\hat{\vec{p}}\hat{H}^{-1}t+\frac{i}{2}(\vec{\alpha}-
\hat{\vec{p}}\hat{H}^{-1})\hat{H}^{-1}\left (e^{-2i\hat H t}-1\right ).
\label{xdirac}
\end{equation}
This equation shows that a free electron surprisingly performs a very 
complicated oscillatory motion. The first two terms are just what is expected,
because
$$\vec{p}H^{-1}=\frac{\vec{p}}{p_0}\frac{H}{p_0}=\frac{\vec{p}}{p_0}\left (
\Lambda_+(\vec{p})-\Lambda_-(\vec{p})\right )$$
is essentially the conventional relativistic velocity for positive-energy
wave functions. But the last term in (\ref{xdirac}) is at odds with our 
classical intuition. According to it the Dirac electron executes rapid 
oscillatory motion, which Schr\"{o}dinger called ``Zitterbewegung''.
As $\vec\alpha(t)\cdot\vec\alpha(t)=1$, the instantaneous value of electron's
velocity during this trembling motion is always 1 (that is the velocity of 
light). The Zitterbewegung is a result of an interference between positive 
and negative frequency Fourier components of the particle wave-packet. The way
the Zitterbewegung shows itself depends on the character of the particle 
wave-packet \cite{75}. In the case of plane wave (not localized particles)
one has a steady-state violent oscillations with amplitude $\sim 1/m$ and
angular frequency $\sim 2m$ ($\hbar/mc$ and $2mc^2/\hbar$ if $\hbar$ and $c$
are restored. For the electron $\hbar/mc=3.85\cdot 10^{-3}$~{\rm \AA} and
$2mc^2/\hbar=1.55\cdot 10^{21}~\mathrm{Hz}$).

For the Newton-Wigner position operator we will have
$$\frac{d\hat{\vec{Q}}(t)}{dt}=i[\hat H,\hat{\vec{Q}}(t)]=
ie^{-iW}e^{i\hat{H}_{(W)}t}[\hat{H}_{(W)},\hat{\vec{Q}}_{(W)}]
e^{-i\hat{H}_{(W)}t}e^{iW}.$$
But in the Foldy-Wouthuesen representation
$$\hat{H}_{(W)}=\beta p_0,\;\; \hat{\vec{Q}}_{(W)}=\frac{1}{2}(1+\beta)
i\frac{\partial}{\partial \vec{p}}$$
and 
$$[\hat{H}_{(W)},\hat{\vec{Q}}_{(W)}]=-\frac{i}{2}(1+\beta)\frac{\vec{p}}
{p_0}.$$
Therefore
$$\frac{d\hat{\vec{Q}}(t)}{dt}=ie^{-iW}\frac{1}{2}(1+\beta)\frac{\vec{p}}
{p_0}\,e^{iW}=\frac{\vec{p}}{p_0}\Lambda_+(\vec{p})$$
and
\begin{equation}
\hat{\vec{Q}}(t)=\hat{\vec{Q}}+\frac{\vec{p}}{p_0}\Lambda_+(\vec{p})t,
\label{QNW}
\end{equation}
where $\hat{\vec{Q}}\equiv \hat{\vec{Q}}(0)$ is the initial value of the
Newton-Wigner position operator given by (\ref{FWmp}).
As we see, the Zitterbewegung is absent in the time development of the 
Heisenberg picture Newton-Wigner position operator. This is not surprising
because only positive frequencies are present in the Fourier components
of wave packets that are localized in the Newton-Wigner sense. However this
absence of trembling has a price: Newton-Wigner wave-packets cannot be 
localized sharper than $1/m$. To see this, let us consider the Newton-Wigner
eigenfunction (\ref{eq7}) in configuration space. 
$$\psi^{(s)}_{(\vec{y})}(\vec{x})=\frac{1}{(2\pi)^3}\int e^{i\vec{p}\cdot 
(\vec{x}-\vec{y})}\sqrt{\frac{2p_0}{p_0+m}}\Lambda_+(\vec{p})\,u(\vec{0},s)\,
d\vec{p}.$$
By using
$$u(\vec{0},s)=\frac{1}{2}(1+\gamma_0)u(\vec{0},s),\;\;
\Lambda_+(\vec{p})\frac{1}{2}(1+\gamma_0)=\frac{1}{2p_0}
\left ( \begin{array}{cc} p_0+m & 0 \\ \vec{p}\cdot\vec{\sigma} & 0  
\end{array} \right ),$$
we get
$$\psi^{(s)}_{(\vec{y})}(\vec{x})=\frac{1}{(2\pi)^3}\int e^{i\vec{p}\cdot
(\vec{x}-\vec{y})}\frac{1}{\sqrt{2p_0(p_0+m)}}\left ( \begin{array}{cc}
p_0+m & 0 \\ \vec{p}\cdot\vec{\sigma} & 0 \end{array} \right )u(\vec{0},s)\,
d\vec{p},$$
or
$$\psi^{(s)}_{(\vec{y})}(\vec{x})=\left ( \begin{array}{cc} A(\vec{z};m) & 0 
\\ -2i\vec{\sigma}\cdot\frac{\partial^2 A(\vec{z};m)}{\partial \vec{z}
\partial m} & 0 \end{array} \right )u(\vec{0},s),$$
where $\vec{z}=\vec{x}-\vec{y}$ and
$$A(\vec{z};m)=\frac{1}{(2\pi)^3}\int e^{i\vec{p}\cdot\vec{z}}\sqrt{\frac{
p_0+m}{2p_0}}\,d\vec{p}.$$
For evaluation of the latter function, it is convenient to decompose
$$\sqrt{1+\frac{m}{p_0}}=1+\sum\limits_{n=1}^\infty a_n \left (\frac{m}{p_0}
\right )^n.$$
Then
\begin{equation}
A(\vec{z};m)=\frac{\delta(\vec{z})}{\sqrt{2}}+\frac{\sqrt{2}}{(2\pi)^2}
\sum\limits_{n=1}^\infty a_n\int\limits_0^\infty p \left (\frac{m}{p_0}
\right )^n \frac{\sin{(pz)}}{z}\, dp,
\label{Az}
\end{equation}
where $p=|\vec{p}|$ and $z=|\vec{z}|$.
But
$$I=m^n \int\limits_0^\infty \frac{p}{p_0^n} \frac{\sin{(pz)}}{z} dp=
-\frac{m^n}{z}\frac{d}{dz}\int\limits_0^\infty \frac{\cos{(pz)}}{(p^2+m^2)^
\frac{n}{2}}\, dp.$$
This latter integral can be evaluated by using \cite{77}
$$K_\nu (mz)=\frac{\Gamma(\nu+\frac{1}{2}) (2m)^\nu}{\sqrt{\pi}z^\nu}
\int\limits_0^\infty \frac{\cos{(pz)}}{(p^2+m^2)^{\nu+\frac{1}{2}}}\,dp$$
and
$$\frac{1}{z}\frac{d}{dz}\left [ z^\nu K_\nu (z)\right ]=-z^{\nu-1}
K_{\nu-1} (z),$$
where $K_\nu (z)$ is the Macdonald function. The result is
$$I=\frac{\sqrt{\pi}m^3}{\Gamma(\frac{n}{2})2^{\frac{n-1}{2}}} (mz)^{\frac
{n-3}{2}} K_{\frac{n-3}{2}} (mz).$$
Substituting this into (\ref{Az}), we get
$$A(\vec{z};m)=\frac{1}{\sqrt{2}}\left [ \delta(\vec{z})+\frac{\sqrt{\pi}m^3}
{(2\pi)^2}\sum\limits_{n=1}^\infty \frac{a_n}{\Gamma(\frac{n}{2})}\left (
\frac{mz}{2}\right )^{\frac{n-3}{2}}K_{\frac{n-3}{2}} (mz)\right ].$$
But for large arguments \cite{77}
$$K_\nu(z)\approx \sqrt{\frac{\pi}{2z}}e^{-z}\left \{1+\frac{4\nu^2-1}
{8z}\right \}.$$
Therefore the function $A(z;m)$ decays for large $|\vec{x}-\vec{y}|$ as 
$e^{-m|\vec{x}-\vec{y}|}$ indicating that the localization provided by the
$\psi^{(s)}_{(\vec{y})}(\vec{x})$ wave-function is no better than $1/m$.

It seems therefore that the quantum mechanics and relativity create conceptual
problems for Zeno's prescription to observe Achilles and the tortoise
race (check the tortoise position when Achilles reach the position tortoise
occupied at previous step). If we use the Dirac position for this goal, we will
face Zitterbewegung, and if we use the Newton-Wigner one, the contenders will 
be not sharply localized. The notion of object's localization, which seems so 
benignly obvious in classical mechanics, undergoes a radical change when 
distances are smaller than the Compton wavelength of the object. At first sight
the notion of localization at Compton scales appears as purely academic concept
because even for electron the Compton wavelength is very small and the 
Zitterbewegung frequency very high, far beyond the present day experimental
accessibility. But this situation can be changed in near future thanks to 
narrow-gap semiconductors \cite{78}.

In such semiconductors the dispersion relation between the energy and the 
wavenumber for electrons is analogous to that of relativistic electrons in 
vacuum and as a result effective relativity arises with  maximum velocity  
of about $10^8~\mathrm{cm/sec}$ instead of the velocity of light \cite{78,79}.
Such systems can be used to model various relativistic phenomena the 
Zitterbewegung included. The amplitude of the corresponding trembling motion 
in semiconductors can be quite large, as much as 64~{\rm \AA} for 
InSb \cite{78}, and can be experimentally observed using high-resolution 
scanning-probe microscopy imaging techniques \cite{80}.

\section{Quantum revivals of Zeno}
Achilles is localized not only at the initial instant but remains so during 
the whole race. Maybe for Zeno this facet of classical world was not 
paradoxical at all but it appears not so trivial at modern times, after the 
quantum revolution. In quantum mechanics wave-packets spread as time goes by
and it is not after all obvious how the classical reality with its definiteness
arises from the weird quantum world. It is assumed, usually, that classical
behaviour of macroscopic objects, like Achilles and tortoise, is something 
obvious and always guaranteed. But this is not correct. For example, a 
cryogenic bar gravity-wave detector must be treated as a quantum harmonic 
oscillator even though it may weigh several tons \cite{81}. Superconductivity
and superfluidity provide another examples of quantum behaviour at 
macroscopic scales.

\begin{figure}[htb] 
  \begin{center}
    \mbox{\epsfig{figure=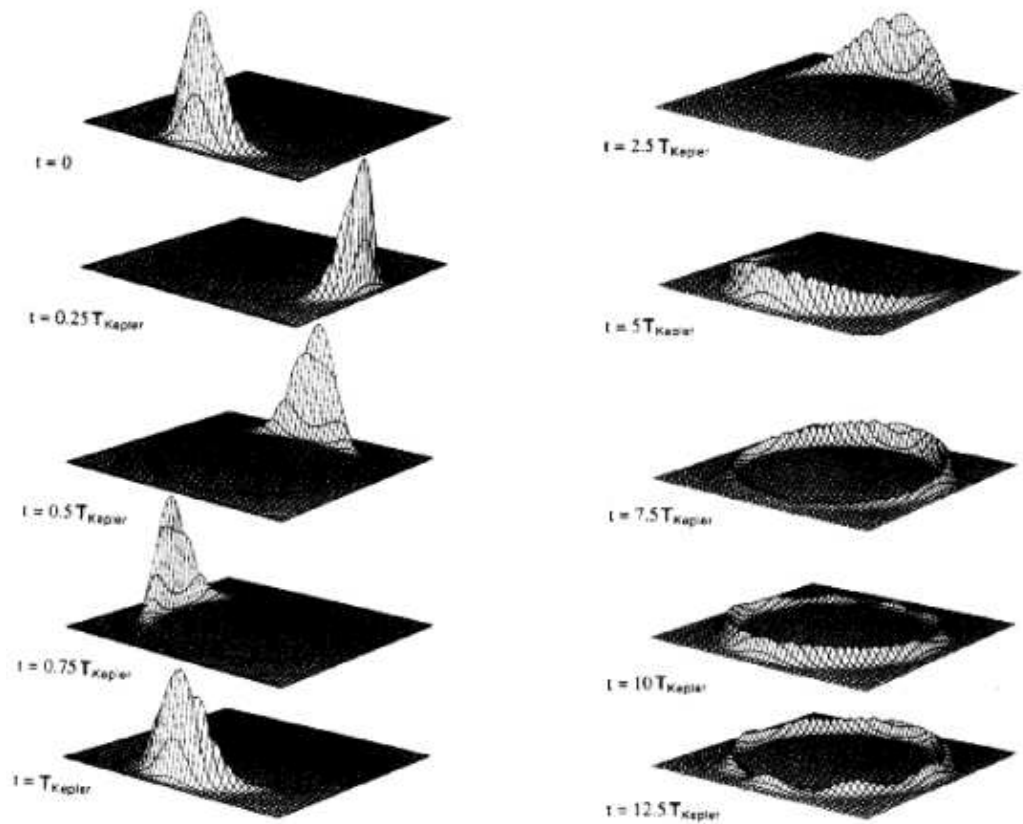}}
  \end{center}
\caption {Circular-orbit wave packet at initial stages of its time evolution
\cite{82}.
}
\label{kepler}
\end{figure}

To illustrate surprises that a quantum particle on a racetrack can offer, let 
us consider a wave packet of an electron in a hydrogen atom constructed from 
a superposition of highly excited stationary states centered at a large 
principal quantum number $\bar n =320$ \cite{82}. Initially the wave packet is
well-localized near a point on the electron's classical circular orbit. Then
it propagates around the orbit in accordance with the correspondence principle.
During the propagation the wave packet spreads along the orbit and after 
a dozen classical periods $T_{Kepler}$ appears as nearly uniformly distributed
around the orbit (Fig.\ref{kepler})

\begin{figure}[htb] 
  \begin{center}
    \mbox{\epsfig{figure=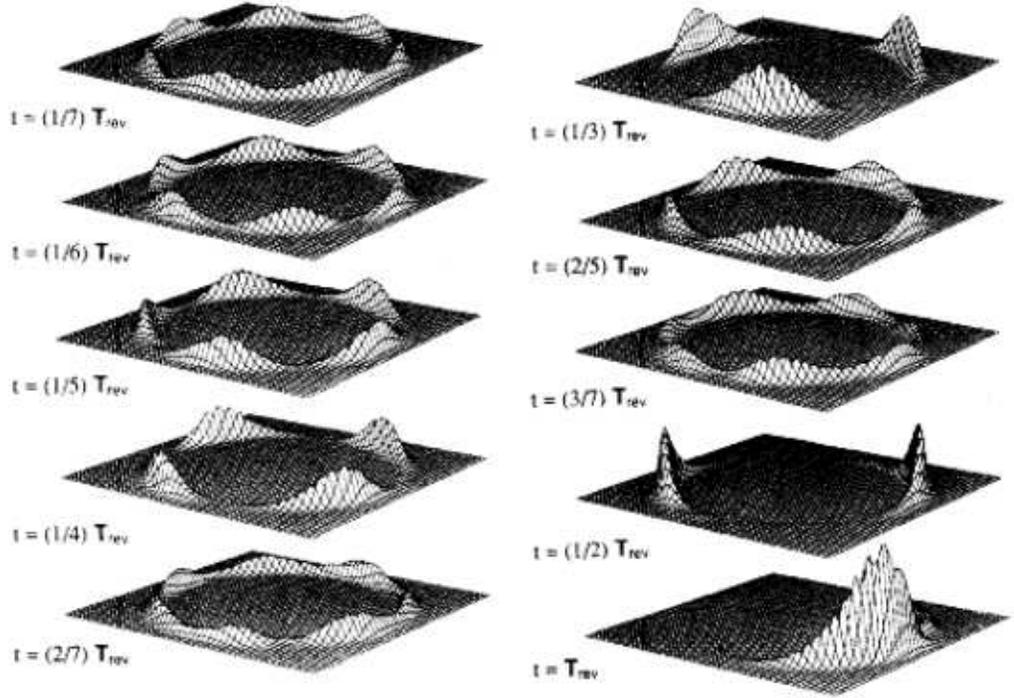}}
  \end{center}
\caption {Wave-packet revival and fractional revivals \cite{82}.}
\label{revival}
\end{figure}

If we wait long enough something extraordinary will happen: after some time
$T_{rev}$ the wave packet contracts and reconstructs its initial form 
(Fig.\ref{revival}). This resurrection of the wave packet from the dead is 
called a quantum revival and it is closely related to the Talbot effect in
optics \cite{83}. In many circumstances the revivals are almost perfect and 
repeat as time goes on. At times $(k_1/k_2)T_{rev}$, where $k_1$ and 
$k_2$ are two mutually prime numbers, fractional revivals happen and the
wave packet consists of several high-correlated smaller clones of the original 
packet.

To explain the origin of quantum revivals, let us consider a particle of 
mass $m$ in an infinite square well of width $L$ \cite{84,85}. The energy 
spectrum of the system is given by
$$E_n=\frac{1}{2m}\left (\frac{\pi\hbar}{L}\right)^2 n^2.$$
Suppose the initial wave function is
$$\psi(x;0)=\sum\limits_{n=1}^\infty a_n \phi_n(x),$$
where $\phi_n(x)$ are the energy eigenfunctions. Time evolution of this wave
function is described by
$$\psi(x;t)=\sum\limits_{n=1}^\infty a_n e^{-\frac{i}{\hbar}E_n t}\phi_n(x).$$
Therefore, if there exists such a revival time $T_{rev}$ that
\begin{equation}
\frac{E_n}{\hbar}T_{rev}=2\pi N_n +\varphi
\label{Trev}
\end{equation}
for all nonzero $a_n$, where  $N_n$ is an integer that can depend on $n$
and $\varphi$ does not depend on $n$, then $\psi(x;T_{rev})$ will describe
exactly the same state as $\psi(x;0)$.

But (\ref{Trev}) is fulfilled if
$$T_{rev}=\frac{4mL^2}{\pi\hbar}$$
and $\varphi=0$. Therefore any quantum state in an infinite square well will
be exactly revived after a time $T_{rev}$. Note that the classical period of
bouncing back and forth between the walls is
$$T_{cl}=\sqrt{\frac{2m}{E}}L$$
and for highly excited states ($n\gg 1$)
$$T_{cl}=\frac{2mL^2}{n\pi\hbar}\ll T_{rev}.$$

Now let us consider the quantum state $\psi(x;t)$ at half of revival time
$$\psi(x;T_{rev}/2)=\sum\limits_{n=1}^\infty a_n e^{-i\pi n^2}\phi_n(x)=
\sum\limits_{n=1}^\infty (-1)^n a_n \phi_n(x),$$
where we have used $e^{-i\pi n^2}=(-1)^n$ identity. But
$$\phi_n(x)=\sqrt{\frac{2}{L}}\sin\frac{n\pi x}{L}=-(-1)^n\phi_n(L-x).$$
Therefore
$$\psi(x;T_{rev}/2)=-\sum\limits_{n=1}^\infty a_n \phi_n(L-x)=
-\psi(L-x;0)$$
and we have the perfect revival of the initial quantum state but at a location 
$L-x$ which mirrors the initial position about the center of the well.

At one quarter of the revival time
$$\psi(x;T_{rev}/4)=\sum\limits_{n=1}^\infty a_n e^{-i\pi n^2/2}\phi_n(x).$$
But 
$$e^{-i\pi n^2/2}=\cos{\frac{n^2\pi}{2}}-i\sin{\frac{n^2\pi}{2}}=
\left \{\begin{array}{c} 1,\;\; {\mathrm if} \;\; n \;\;{\mathrm even} \\ 
-i,\;\; {\mathrm if}\;\; n \;\;{\mathrm odd} \end{array} \right .$$
Therefore
$$\psi(x;T_{rev}/4)=\sum\limits_{n\;even} a_n \phi_n(x)-i
\sum\limits_{n\;odd} a_n \phi_n(x).$$
Comparing this expression to
$$\psi(x;0)=\sum\limits_{n\;even} a_n \phi_n(x)+
\sum\limits_{n\;odd} a_n \phi_n(x)$$
and
$$\psi(L-x;0)=-\sum\limits_{n\;even} a_n \phi_n(x)+
\sum\limits_{n\;odd} a_n \phi_n(x),$$
we deduce that \cite{85}
$$\psi(x;T_{rev}/4)=\frac{1-i}{2}\psi(x;0)-\frac{1+i}{2}\psi(L-x;0).$$
Therefore we have the perfect fractional revival at a time $T_{rev}/4$
when two smaller copies of the initial wave packet appear at locations
$x$ and $L-x$. 

\begin{figure}[htb] 
  \begin{center}
    \mbox{\epsfig{figure=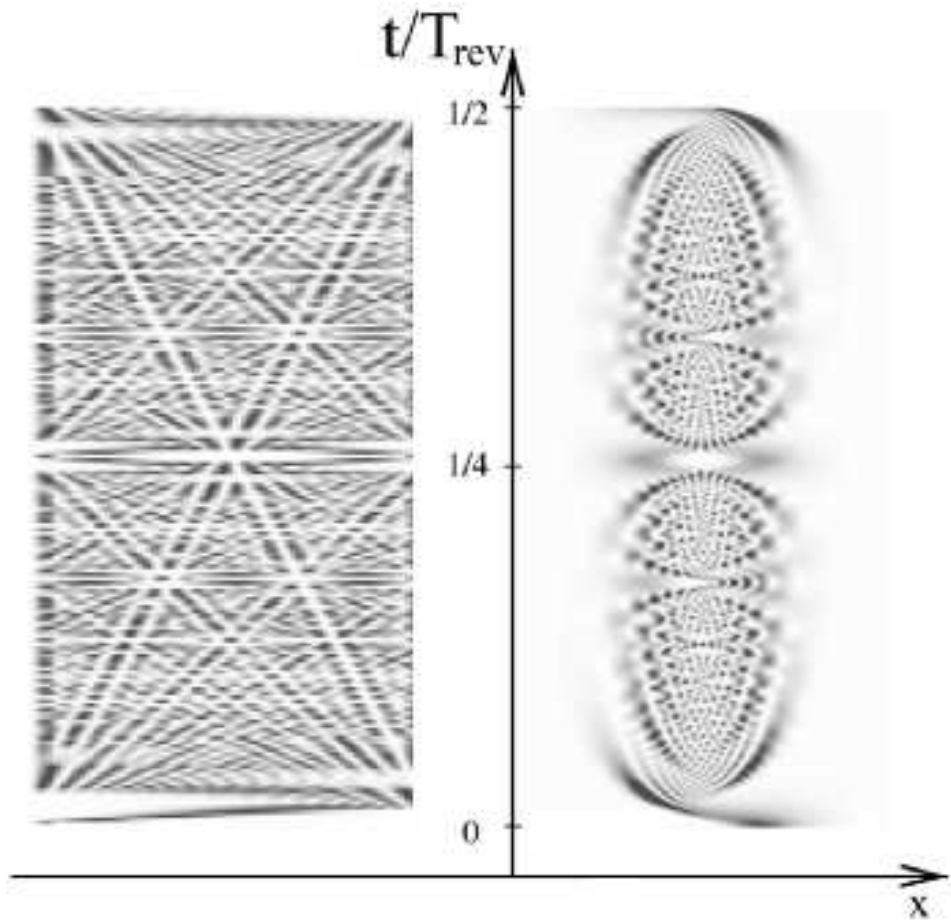}}
  \end{center}
\caption {The quantum carpets for the P\"{o}schel-Teller and Rosen-Morse 
potentials \cite{85}.}
\label{carpet}
\end{figure}

Space-time structure of the probability density $|\psi(x;t)|^2$ is also
very interesting. When plotted over long time periods (of the order of
$T_{rev}$) it exhibits fine interference patterns  known as quantum carpets
\cite{86,87}. Some examples are shown in Fig.\ref{carpet}. 

Contemporary experimental technique allows to investigate quantum revivals and
carpets and many theoretical results have been confirmed by experiments 
\cite{83,88}. Of course, nothing remotely similar to this weird phenomena
happens during the Achilles and the tortoise race. And we come to one more
paradox: why Zeno's paradoxes, formulated purely in classical terms, make 
complete sense for us? The answer may sound like this \cite{81}: ``The 
environment surrounding a quantum system can, in effect, monitor some of the 
systems observables. As a result, the eigenstates of these observables 
continuously decohere and can behave like classical states''. Objects have no
a priori classical properties. These properties are emergent phenomenon and
come into being only through the very weak interaction with the ubiquitous 
degrees of freedom of the environment. Amazingly, the emergence of the 
classical world seems to be just another side of the quantum Zeno effect.
As an example one can consider a large chiral molecule (like sugar) which can 
have both left-handed and right-handed classical spatial structures 
\cite{89,90}. For symmetry reasons, the ground state is equal mixture of both 
chiral states. Chiral molecules are never found in energy eigenstates and this 
is probably not surprising because such states are examples of Schr\"{o}dinger 
cat states (like a superposition of a dead and an alive cat) which look truly 
absurd from classical viewpoint. But the real reason why nonclassical states
of the Chiral molecules are not observed is that it is chirality (not parity)
that is recognized by the environment, for example by scattered air molecules.
The chirality of the molecule is thus continuously ``observed'' by the 
environment and therefore cannot change because of quantum Zeno effect.

\section{Zeno meets quantum gravity}
Another twist to the story of localization is added when one tries to
incorporate gravity into a quantum theory. Any sharp localization of the
system creates a significant local energy density due to uncertainty
relations and therefore changes the space-time metric according to the
philosophy of general relativity. This can effect another localization
effort nearby in an unavoidable manner \cite{91} and as a result it will 
matter whether x-position measurement is carried before or after y-position
measurement. Moreover, localization sharper than the Plankian scale creates a
singularity in the space-time metric and therefore is problematic. We expect
consequently that any coherent quantum gravity theory will not only bring the 
fundamental Plankian scale as the limit of space-time divisibility with it, 
but also a non-commutative space-time geometry.

Not only the concept of localization of material objects but also operational 
meaning of the space-time itself is expected to be lost at Plankian scales.
The principles of quantum mechanics and general relativity limit the accuracy
of space-time distance measurements. The argument goes as follows \cite{92}.

Suppose we want to measure the initial distance between Achilles and the 
tortoise. We can attach a small mirror to the tortoise while a clock with 
light-emitter and receiver to Achilles. When the clock reads zero a light
signal is sent to be reflected by the mirror. If the reflected signal arrives
back at time $t$ then the distance is $l=ct/2$. Note that this is a quite
realistic scheme used, for example, in Lunar Laser Ranging experiments 
\cite{93}. Quantum mechanics sets some limits on ultimate precision which
can be reached in such distance measurements. Let, for example, the initial
uncertainty in the clock's position be $\Delta x$. Then according to 
uncertainty relation its speed is also uncertain with the spread
$$\Delta v\ge \frac{1}{2}\frac{\hbar}{m\Delta x}.$$
where $m$ is the mass of the clock. At time $t$ the uncertainty of the 
clock's position will be larger
$$\Delta x(t)= \Delta(x(0)+vt)=\sqrt{(\Delta x)^2+(t\Delta v)^2}=
\sqrt{(\Delta x)^2+\frac{1}{4}\frac{t^2\hbar^2}{m^2(\Delta x)^2}}.$$
The optimal uncertainty at initial time that minimizes the uncertainty at time
$t$ is given by
$$(\Delta x)^2=\frac{1}{2}\frac{t\hbar}{m}.$$
Therefore the minimal uncertainty at time $t$ is
$$\Delta x(t)=\sqrt{\frac{t\hbar}{m}}=\sqrt{\frac{2l\hbar}{mc}}
=\lambda_C\sqrt{\frac{2l}{\lambda_C}},$$
where $\lambda_C=\hbar/mc$ is the Compton wavelength of the clock. The mirror
contributes similarly to the overall uncertainty of the distance $l$ to be 
measured. Therefore, ignoring small factors of the order of 2, we obtain an
order of magnitude estimate \cite{92,94} (the argument actually goes back to 
Wigner \cite{95,96})
\begin{equation}
\Delta l \ge \lambda_C\sqrt{\frac{l}{\lambda_C}}.
\label{Deltal}
\end{equation}
Hence we need massive clock to reduce the uncertainty. But the mass of the 
clock cannot be increased indefinitely because the distance $l$ should be
greater than the clock's Schwarzschild radius. Otherwise the clock will 
collapse into a black hole as it is assumed that its size is smaller than $l$. 
Therefore 
$$l\ge \frac{Gm}{c^2}$$
and inserting this into (\ref{Deltal}) we get
\begin{equation}
\Delta l \ge l_P,
\label{DeltaP}
\end{equation}
where
$$l_P=\sqrt{\frac{\hbar G}{c^3}}$$
is the Planck length.

Ng and Van Dam argue even for a more restrictive bound \cite{92}. suppose
the clock consists of two parallel mirrors a distance $d$ apart. Then its one 
tick cannot be less than $d/c$ and this implies $\Delta l \ge d$ if the clock
is used for timing in distance measurements. But $d$ should be greater than 
the clock's Schwarzschild radius. Therefore
\begin{equation}
\Delta l \ge \frac{Gm}{c^2}. 
\label{DeltaG}
\end{equation}
Squaring (\ref{Deltal}) and multiplying the result by (\ref{DeltaG}), we 
eliminate the clock's mass and obtain $(\Delta l)^3\ge  l l_P^2$. Therefore
\begin{equation}
\Delta l \ge l_P\left (\frac{l}{l_P}\right)^{1/3}. 
\label{DeltalP}
\end{equation}

It seems there is a common consensus about the validity of (\ref{DeltaP}).
While the more stronger bound (\ref{DeltalP}) is still under debate (see,
for example, \cite{94}). Nevertheless this latter bound is consistent with the 
holographic principle \cite{97} which states that the information content of
any region of space cannot exceed its surface area in Planck units. Indeed
\cite{98}, suppose we have a cube of dimension $l\times l\times l$ and every
cell $\Delta l\times \Delta l\times \Delta l$ of this cube can be used to 
store one bit of information. $\Delta l$ cannot be made less than dictated
by the bound (\ref{DeltalP}), otherwise it will be possible to measure the 
size of the cube in $\Delta l$ units and reach the precision superior to the 
limit (\ref{DeltalP}). Therefore the information content of the cube 
$$N=\frac{l^3}{(\Delta l)^3}\le \frac{l^2}{l_P^2}.$$

In any case, we can conclude that basic principles of quantum mechanics and
general relativity strongly suggest discreteness of space-time at some
fundamental scale. Zeno anticipated such possibility and attacked it with
another couple of paradoxes. The first one, the Arrow, states that \cite{2}
\begin{itemize}
\item
{\it If everything is either at rest or moving when it occupies a space equal 
to itself, while the object moved is always in the instant, a moving arrow is 
unmoved.}
\end{itemize}
Stated differently, if a particle exists only at a sequence of discrete 
instants of time, what is the instantaneous physical properties of a moving 
particle which distinguishes it from the not moving one? And if there is no
such properties (well, the notion of instantaneous velocity requires the 
concept of limit and thus is inappropriate in discrete space-time) how the
motion is possible?

Modern physics changed our perspective of particles and motion and the Arrow
seems not so disturbing today. For example we can defy it by stating that the
arrow cannot be at rest at definite position according to the uncertainty
principle. Alternatively, we can evoke special relativity and say that there
is a difference how the world looks for a moving arrow and for an arrow at
rest \cite{2}: they have different planes of simultaneity. Special relativity
can cope also with the second Zeno paradox against space-time discreteness,
the Stadium: 
\begin{itemize}
\item
{\it Consider two rows of bodies, each composed of an equal number of bodies 
of equal size.  They pass each other as they travel with equal velocity in 
opposite directions.  Thus, half a time is equal to the whole time.}
\end{itemize}
If motion takes place in discrete quantum jumps then there should be an 
absolute upper bound on velocity. The maximum velocity is achieved
then all jumps are in the same direction. The Stadium is intended to show
logical impossibility of the maximum relative velocity. Suppose the rows
of bodies from the paradox move at maximum or nearly maximum speed. Then 
in the rest frame of the first row other bodies are approaching  at twice or 
so the maximum possible speed. Now we know that the latter  inference is
not sound. But Zeno was right that radical change of classical concepts of 
space and time is necessary to assimilate an  observer-independent maximum
velocity. ``Space by itself and time by itself are to sink fully into shadows 
and only a kind of union of the two should yet preserve autonomy'' - to quote
Minkowski from his famous Cologne lecture in 1908.

In fact space-time discreteness coupled with the relativity principle assumes
two invariant scales: not only the maximum velocity but also the minimum 
length. In this respect the special relativity refutation of Zeno is not 
complete. Only recently a significant effort was invested in developing
Doubly Special Relativity \cite{99,100} -  relativity theories with two 
observer-independent scales. These developments, although interesting, still
are lacking experimental confirmation.

Doubly Special Relativity, if correct, should be a limiting case of quantum
gravity - an ultimate theory of quantum space-time and a major challenge of
contemporary physics to combine general relativity with quantum mechanics.
This synthesis is not achieved yet but curiously enough its outcome may
turn out to vindicate Parmenidean view that time and change are not 
fundamental reality. In any case the problem of time seems to be central in 
quantum gravity because time plays conceptually different roles in quantum
mechanics and general relativity not easy to reconcile \cite{101,102,103,104,
105}.

In quantum mechanics time is an external parameter, not an observable in the 
usual sense -- it is not represented by an operator. Indeed, suppose there is
an operator $\hat T$, representing a perfect clock, such that in the 
Heisenberg picture $\hat T(t)=e^{i\hat Ht}\hat T e^{-i\hat Ht}=t$. Then
(we are using again $\hbar=1$ convention)
$$i[\hat H,\hat T(t)]=\frac{d}{dt}\hat T(t)=1.$$
Therefore  $[\hat T(t),\hat H]=i$ and $[\hat H,e^{i\alpha\hat T}]=\alpha 
e^{i\alpha\hat T}$. If $\psi$ is an energy eigenstate with energy $E$ then
$$\hat H e^{i\alpha\hat T}\psi=\left \{ [\hat H, e^{i\alpha\hat T}]+
e^{i\alpha\hat T} \hat H\right \}\psi=(E+\alpha)e^{i\alpha\hat T}\psi.$$
Therefore $e^{i\alpha\hat T}\psi$ is also the energy eigenstate with energy
$E+\alpha$ and the Hamiltonian spectrum cannot be bounded from below as it 
usually is. This argument goes back to Pauli \cite{106}. Alternatively we can
resort to the Stone-von Neumann theorem \cite{107} and argue that the 
canonically conjugate $\hat T$ and $\hat H$ are just the disguised versions of 
the position and momentum operators and therefore must have unbounded spectra.

However, there are subtleties in both the Pauli's argument \cite{106} 
as well as
in the Stone-von Neumann theorem \cite{108} and various time operators are
suggested occasionally. Note that many useful concepts in physics are 
ambiguous or even incorrect from the mathematical point of view. In an 
extremal manner this viewpoint was expressed by Dieudonn\'{e} \cite{109}: 
"When one gets to the mathematical theories which are at the basis of quantum 
mechanics, one realizes that the attitude of certain physicists in the 
handling of these theories truly borders on the delirium''. Of course, this
is an exaggeration but sometimes mathematical refinement leads to new physical
insights and it is not excluded that the last word about the time operator is
not said yet. In any case, subtle is the time in quantum mechanics!

The idea of an event happening at a given time plays a crucial role in quantum
theory \cite{110} and at first sight introduces unsurmountable difference
between space and time. For example \cite{101}, if $\Psi(\vec{x},t)$ is a 
normalized wave function then $\int |\Psi(\vec{x},t)|^2 d\vec{x}=1$ for all 
times, because the particle must be somewhere in space at any given instant of 
time. While $\int |\Psi(\vec{x},t)|^2 dt$ can fluctuate wildly for various 
points of space.

Nevertheless, the conceptual foundation of quantum theory is compatible with
special relativity. Absolute Newtonian time is simply replaced by Minkowski 
spacetime fixed background and unitary representations of the Poincar\'{e}
group can be used to develop quantum theory. Situation changes dramatically
in general relativity. ``There is hardly any common ground between the general
theory of relativity and quantum mechanics'' \cite{95}. The central problem is
that spacetime itself becomes a dynamical object in general relativity. Not
only matter is influenced by the structure of spacetime but the metric
structure of spacetime depends on the state of ambient matter. As a result,
the spatial coordinate $\vec{x}$ and the temporal coordinate $t$ lose any
physical meaning whatsoever in general relativity. If in 
non-general-relativistic physics (including special relativity) the 
coordinates correspond to readings on rods and clocks, in general relativity
they correspond to nothing at all and are only auxiliary quantities which
can be given arbitrary values for every event \cite{95,111}. 

Already at classical level, general relativity is a great deal Parmenidean
and usually some tacit assumptions which fix the coordinate system is needed
to talk meaningfully about time and time-evolution. To quote Wigner \cite{95}
``Evidently, the usual statements about future positions of particles, as
specified by their coordinates, are not meaningful statements in general
relativity. This is a point which cannot be emphasized strongly enough and is 
the basis of a much deeper dilemma than the more technical question of the
Lorentz invariance of the quantum field equations. It pervades all the general
theory, and to some degree we mislead both our students and ourselves when we
calculate, for instance, the mercury perihelion motion without explaining how
our coordinate system is fixed in space, what defines it in such a way that it
cannot be rotated, by a few seconds a year, to follow the perihelion's 
apparent motion''.

In quantum theory the situation only worsens. The Hamiltonian generates the 
time evolution of quantum system. But the equations of motion of general 
relativity are invariant under time reparametrization. Therefore the time 
evolution is in fact unobservable - it is a gauge. The Hamiltonian vanishes
and the Schr\"{o}dinger equation in cosmology - the Wheeler-DeWitt equation
does not contain time. ``General relativity does not describe evolution in 
time: it describes the relative evolution of many variables with respect to
each other'' \cite{111}. Zeno and Parmenides with their strange idea that
time and change are some kind of illusion still have chance in quantum 
gravity!

\section{Conclusion} 
The main conclusion of this paper is that physics is beautiful. Questions 
aroused two and half millennium ago and scrutinized many times are still not
exhausted. Zeno's paradoxes deal with fundamental aspects of reality like
localization, motion, space and time. New and unexpected facets of these 
notions come into sight from time to time and every century finds it 
worthwhile to return to Zeno over and over. The process of approaching to 
the ultimate resolution of Zeno's paradoxes seems endless and our 
understanding of the surrounding world is still incomplete and fragmentary.  

``Nevertheless, I believe that there is something great in astronomy, in 
physics, in all the natural sciences that allows the human being to look 
beyond its present place and to arrive at some understanding of what goes on 
beyond the insignificant meanness of spirit that so often pervades our 
existence. There is a Nature; there is a Cosmos; and we walk towards the 
understanding of it all. Is it not wonderful? There are many charms in the 
profession; as many charms as in love provided, of course, that they are not 
in the service of mercantile aims'' \cite{112}.

\end{document}